\documentclass[amsmath,amssymb,showpacs,aps,prd,superscriptaddress,twocolumn,floatfix,nofootinbib]{revtex4}

\usepackage{amsmath, amssymb, epsfig, color}

\newcommand{\LW}{{ Li\'{e}nard-Wiechert }}
\newcommand{\etal}{{\it et al.}}

\newcommand{\tp}{t^{\prime}}
\newcommand{\dt}{\mathrm{d}t}
\newcommand{\dtp}{\mathrm{d}t^{\prime}}
\newcommand{\tpo}{t^{\prime}_0}

\newcommand{\deltp}{\Delta t^{\prime}}
\newcommand{\dtpp}{\mathrm{d}t^{\prime \prime}}
\newcommand{\tpps}{ t^{\prime \prime 2} }
\newcommand{\tpp}{t^{\prime \prime}}

\DeclareMathAlphabet{\mathpzc}{OT1}{pzc}{m}{it}

\begin{document}

\title{General description of electromagnetic radiation processes based on instantaneous charge acceleration in `endpoints'}

\author{Clancy W. James\footnote{corresponding author: clancy.james@physik.uni-erlangen.de}}
\affiliation{Department of Astrophysics, IMAPP, Radboud University Nijmegen, P.O.Box 9010, 6500GL Nijmegen, The Netherlands}
\affiliation{Friedrich-Alexander-Universit\"at Erlangen-N\"urnberg, Erlangen Centre for Astroparticle Physics, Erwin-Rommel-Str. 1, 91058 Erlangen, Germany}
\author{Heino Falcke}
\affiliation{Department of Astrophysics, IMAPP, Radboud University Nijmegen, P.O.Box 9010, 6500GL Nijmegen, The Netherlands}
\affiliation{ASTRON, Postbus 2, NL-79900AA Dwingeloo, the Netherlands}
\author{Tim Huege\footnote{corresponding author: tim.huege@kit.edu}}
\affiliation{Karlsruher Institut f\"{u}r Technologie, Institut f\"{u}r Kernphysik - Campus Nord, 76021 Karlsruhe, Germany}
\author{Marianne Ludwig}
\affiliation{Karlsruher Institut f\"{u}r Technologie, Institut f\"{u}r Experimentelle Kernphysik - Campus S\"{u}d, 76128 Karlsruhe, Germany}


\begin{abstract}
We present a new methodology for calculating the electromagnetic radiation from accelerated charged particles. Our formulation --- the `endpoint formulation' --- combines numerous results developed in the literature in relation to radiation arising from particle acceleration using a complete, and completely general, treatment. We do this by describing particle motion via a series of discrete, instantaneous acceleration events, or `endpoints', with each such event being treated as a source of emission. This method implicitly allows for particle creation/destruction, and is suited to direct numerical implementation in either the time- or frequency-domains. In this paper, we demonstrate the complete generality of our method for calculating the radiated field from charged particle acceleration, and show how it reduces to the classical named radiation processes such as synchrotron, Tamm's description of Vavilov-Cherenkov, and transition radiation under appropriate limits. Using this formulation, we are immediately able to answer outstanding questions regarding the phenomenology of radio emission from ultra-high-energy particle interactions in both the Earth's atmosphere and the Moon. In particular, our formulation makes it apparent that the dominant emission component of the Askaryan Effect (coherent radio-wave radiation from high-energy particle cascades in dense media) comes from coherent `bremsstrahlung' from particle acceleration, rather than coherent Vavilov-Cherenkov radiation.
\end{abstract}

\maketitle

\section{Introduction}

Electromagnetic radiation arising from charged particle motion at scales outside the quantum-mechanical limit can be described completely from Maxwell's equations and the distribution of charges and their accelerations \footnote{For a review of radiating, non-accelerated systems, see Ginzburg \cite{Ginzburg82}. Our treatment will also allow for `apparent acceleration', e.g.\ in the case of transition radiation, and that from time-varying currents, which can be described in terms of charge acceleration.}. Most university courses begin with these fundamental relations, proceed to intermediate results such as the \LW potentials and/or the Larmor formula for the power radiated from an accelerated charge, and use these to derive the properties of classical named radiation processes such as synchrotron, transition, and Vavilov-Cherenkov radiation where an (at least semi-) analytic solution can be found. This approach leads to a greater understanding of both radiation phenomenology, such as relativistic beaming, and of those physical situations where the rather special assumptions on the particle motion required to find an analytic solution apply.

The focus on classical named radiation processes however can leave the impression that these are the fundamental mechanisms of electromagnetically radiating systems, whereas they are really short-hand for a type of charged particle acceleration which results in a particular radiation field. Because of this focus, there is a tendency among physicists to ascribe the radiation from complex systems to a combination of these classical processes. This tendency can lead to confusion however even when a physical situation deviates only slightly from the classical idealised cases. For instance, consider the titles of the following papers: ``Synchrotron radiation of charged particles passing through the boundary between two media'' \cite{Sokolov77}, ``On the \v{C}erenkov threshold associated with synchrotron radiation in a dielectric medium'' \cite{Pogorzelski74}, and ``\v{C}erenkov radiation from an Electron Traveling in a Circle through a Dielectric Medium'' \cite{Erteza62}. The titles could equally have referred to `transition radiation', `Vavilov-Cherenkov radiation', and `synchrotron radiation' respectively. Indeed, numerous papers exist which note that the same fundamental physics can explain multiple mechanisms, for instance Schwinger, Tsai and Erber's ``Classical and Quantum Theory of Synergic Synchrotron-\v{C}erenkov Radiation'' \cite{SchwingerTsaiErber}. However, the problem of which mechanism to attribute to what process can be avoided entirely by going `back to basics' and formulating a general description of radiation processes according to the well-known mantra ``electromagnetic radiation comes from accelerated charges''.

The goal of this paper is therefore to develop a new methodology --- the `endpoint formulation' --- by which the calculation of the radiated fields from accelerated particles can be performed completely generally via a method which is intuitively understandable at an undergraduate level. For generality, our formulation must be equally suited to that of a simple system allowing an analytic solution as to a complex one requiring a numerical solution, and allow not just the calculation of the total radiated power, but also the time- and frequency-dependent electric field strengths.

We thus proceed as follows. In Section \ref{endpoint_derivation}, we begin with expressions for the electric field from the well-known \LW potentials, and use these to develop our formulation, which is based on the radiation from the instantaneous acceleration of a particle from/to rest (an `endpoint'). Given the radiation from a single endpoint, we describe how to use this to calculate the radiation from an arbitrary complex physical situation. In Section \ref{compare}, we use the endpoint formulation in specific applications to numerically reproduce the well-known results from idealised classical phenomena such as synchrotron and transition radiation, and Tamm's description of Vavilov-Cherenkov radiation from finite particle tracks\footnote{As will be discussed later, this radiation is of a different nature to that produced by the Frank-Tamm description of `true' Vavilov-Cherenkov radiation arising from the particle motion, and comes from the `bremsstrahlung' from the implied acceleration events --- or endpoints --- at the ends of each track.}.
Section \ref{discussion} discusses the use and range of applicability of our endpoint formulation. In Section \ref{applications}, we return to the original motivation for this work from the point of view of the authors, and demonstrate how this formulation resolves outstanding questions relating to the calculation of radiation from high-energy particles cascades in the Moon and the Earth's atmosphere. The explanations also serve to illustrate how the application of classical radiation emission mechanisms to complex physical situations has led to incorrect and misunderstood conclusions.

\section{An Endpoint Formulation}
\label{endpoint_derivation}

We wish to describe electromagnetic radiation in terms of particle acceleration. Rather than writing down a general function for the charged particle distribution and its time derivatives, and then deriving results for specific cases of that charge distribution, here we adopt a `bottom-up' approach. Thus we begin by describing the radiation from a simple single radiating unit, that being the instantaneous acceleration of a charged particle either to or from rest, and later we will show how to combine such basic units into more complex physical situations.

Radiation from an instantaneous particle acceleration is known best through Larmor's formula and its relativistic generalisation, which gives the total power radiated per unit frequency and solid angle. In the interest of clarity, and because we wish to preserve phase information, we proceed below to re-derive the emitted radiation in terms of the electric-field in the time- and frequency-domains. By beginning with the \LW potentials only, we hope to emphasise the generality of our result.

\subsection{The \LW approach}

Electric field components due to particle motion and acceleration can be readily separated
using the \LW potentials, which are derived directly from Maxwell's equations in the relativistic case\footnote{For a derivation for $n=1$, see e.g.\ Jackson \cite{Jackson}. Using an arbitrary $n$ produces the result for Eq.\ \ref{lweqn1} from Zas, Halzen, and Stanev \cite{ZHS92} in the case where the relative permeability, $\mu_r$, is unity.}, and reproduced below:
\begin{eqnarray}
\Phi(\vec{x},t) & = & \left[ \frac{e}{(1 - n \vec{\beta} \cdot \hat{r}) R} \right]_{\rm ret} \nonumber \\
\vec{A}(\vec{x},t) & = & \left[ \frac{e \vec{\beta}}{(1 - n \vec{\beta} \cdot \hat{r})} \right]_{\rm ret} \label{lwpot}
\end{eqnarray}
where $R$ is the distance from the point of emission to an observer, $\hat{r}$ a unit vector in the direction of the observer, $\vec{\beta} = \vec{v}/c$ ($\vec{v}$ is the velocity vector of the particle), and $n$ is the medium refractive index. The subscript `ret' denotes evaluation at the retarded time $t^{\prime} = t - n R/c$. Using Eq.\ \ref{lwpot}, it is possible to calculate the total static and vector potentials from a distribution of source charges, by summing the contributions from individual charges in the distribution. These can then be used to calculate the corresponding electric and magnetic fields. This alternative approach is used for instance by Alvarez-Mu\~niz et al.\ in the `ZHAireS' code \cite{zhaires}.  In our methodology however, we calculate the electric fields directly, since from the \LW potentials, the electric field in a dielectric, non-magnetic medium due to a particle of charge $q$ (in c.g.s.\ units) can be expressed (see e.g.\ Jackson \cite{Jackson}) as follows:
\begin{eqnarray}
\vec{E}(\vec{x},t) & = & q \left[ \frac{\hat{r} - n \vec{\beta}}{\gamma^2 (1 - n \vec{\beta} \cdot \hat{r})^3 R^2} \right]_{\rm ret} \nonumber \\
& + & \frac{q}{c} \left[ \frac{\hat{r} \times [(\hat{r} - n \vec{\beta}) \times \dot{\vec{\beta}}]}{(1 - n \vec{\beta} \cdot \hat{r})^3 R} \right]_{\rm ret} \label{lweqn1}
\end{eqnarray}
with $\dot{\vec{\beta}}$ the time-derivative of $\vec{\beta}$, and $\gamma$ the usual relativistic factor of $(1 - \beta^2)^{-0.5}$. The first term is the near-field term, since the strength of the resulting fields falls as $R^{-2}$ --- in the case of $\beta=0$, it reduces to Coulomb's Law. The second term is the radiation term, with the familiar $1/R$ dependence. The well-known maxim `radiation comes from accelerated charges' is seen easily by the dependence of this term on $\dot{\vec{\beta}}$.

In most practical applications, the near-field term presents only a minor correction to the observed fields. From here on we formulate our methodology purely from the radiated field term only. Thus the following expressions for the electric fields will be those arising solely from the particle acceleration. The applicability of this approximation is discussed in Sec.\ \ref{discussion}.

\subsection{Radiation from an endpoint}

The most simple acceleration event is the instantaneous acceleration of a particle at rest at time $\tp=\tpo$ to a velocity $\vec{\beta}=\vec{\beta}^*$, i.e.\ $\dot{\vec{\beta}} = \vec{\beta}^* \delta(\tp-\tpo)$, or equivalently the deceleration of such a particle from velocity $\vec{\beta}=\vec{\beta}^*$ to rest, i.e.\ $\dot{\vec{\beta}} = - \vec{\beta}^* \delta(\tp-\tpo)$. Such events can be termed, respectively, `starting points' and `stopping points', `acceleration' and `deceleration', or `creation' and `destruction' events. We define the electric field resulting from these events as $\vec{E}_{\pm}$, where the acceleration vector $\dot{\vec{\beta}}$ can be either parallel ($+$) or anti-parallel ($-$) to the velocity vector $\vec{\beta}$, corresponding respectively to acceleration (at a starting point) or deceleration (at a stopping point). Since $\vec{\beta}$ changes only in magnitude, we write $\vec{\beta} = \beta \hat{\beta}$ ($\hat{\beta}$ a unit vector), and use similar notation for $\vec{\beta}^* = \beta^* \hat{\beta}$ and the time-derivatives: $\dot{\vec{\beta}} = \dot{\beta} \hat{\beta}$. Thus only the scalar components will need to be expressed as functions of time.

We proceed to derive $\vec{E}_{\pm}$ from the RHS of Eq.\ \ref{lweqn2} in terms of the `lab-time' (observer time) $t$ in both the time- and frequency-domains. Similar derivations in both domains in the case of linear particle tracks (effectively two endpoints --- see Sec.\ \ref{cherenkov}) appear also in Alvarez-Mu\~{n}iz, Romero-Wolf, and Zas \cite{AMRWZ2010}, for the case of ``\v{C}erenkov radiation'': note that in the following no assumption on the nature of the radiation need to be made.

\subsubsection{Frequency-domain derivation}

The expression for the radiated component of the electric field (from Eq.\ \ref{lweqn1}) for the instantaneous particle acceleration described above is:
\begin{eqnarray}
\vec{E}_{\pm}(\vec{x},t) & = &  \pm \frac{q}{c} \left[ \frac{\hat{r} \times [ \hat{r} \times \vec{\beta}^* \delta(\tp-\tpo)]}{(1 - n \vec{\beta} \cdot \hat{r})^3 R} \right]_{\rm ret} \label{lweqn2}
\end{eqnarray}
where we have removed the $\vec{\beta} \times \dot{\vec{\beta}}$ term from Eq.\ \ref{lweqn1} since $\vec{\beta} \parallel \dot{\vec{\beta}}$. We begin the frequency-domain derivation by taking the Fourier-transform of Eq.\ \ref{lweqn2} converted to the retarded time $\tp$ using $t = \tp + R n/c$ and $\dt = \dtp (1 - n \vec{\beta} \cdot \hat{r})$:
\begin{eqnarray}
\vec{E}_{\pm}(\vec{x},\nu)  \equiv \int \dt \, \vec{E}_{\pm}(\vec{x},t) \, e^{2 \pi i \nu t} \nonumber \\
= \int \dtp \, \vec{E}_{\pm}(\vec{x},t(\tp)) (1 - n \vec{\beta} \cdot \hat{r}) \, e^{2 \pi i \nu (\tp + R n/c )}.
\end{eqnarray}
A conceptual and mathematical difficulty to overcome is that at the time of acceleration, $\beta$, and hence $\vec{E}_{\pm}(\vec{x},\nu)$, is undefined. This can be dealt with by letting the acceleration last a finite (but small) time interval $\deltp$, then taking the limit as $\deltp \to 0$. Writing $\tpp = \tp-\tpo$, the acceleration takes place over the interval $0 < \tpp < \deltp$, during which we have $\beta(\tpp) = \beta^* \tpp/\Delta \tp$, $\dot{\beta} = \beta^*/\Delta \tp $, and $ R(\tpp) = R(t^{\prime}_0) - 0.5 c (\beta^*/\Delta \tp) \tpps \hat{\beta} \cdot \hat{r}$. Thus the frequency-domain integral becomes:
\begin{eqnarray}
\vec{E}_{\pm}(\vec{x},\nu) 
 = \pm \lim_{\Delta \tp \rightarrow 0} \frac{q}{c} \, e^{2 \pi i \nu \tpo}\cdot \nonumber \\
 \int_{0}^{\Delta \tp} \hspace{-1mm} \, \frac{\frac{1}{\Delta \tp} e^{2 \pi i \nu (\tpp + n R(\tpp)/c)} }{(1 - \frac{n \tpp}{\Delta \tp} \vec{\beta}^* \cdot \hat{r})^2 R(\tpp)} \left( \hat{r} \times [\hat{r} \times \vec{\beta}^*] \right) \, \dtpp
\end{eqnarray}
This somewhat difficult integral can be greatly simplified by applying the limit $\Delta \tp \to 0$, in which case the integral and limit eventually evaluate to the rather simple form:
\begin{eqnarray}
\vec{E}_{\pm}(\vec{x},\nu) & = & \pm \frac{q }{c} \, \frac{e^{i k R(t^{\prime}_0)}}{R(t^{\prime}_0)} \, \frac{e^{2 \pi i \nu \tpo} }{1- n \vec{\beta}^* \cdot \hat{r}} \, \hat{r} \times [\hat{r} \times \vec{\beta}^*] \label{endpoint_eqn_f}
\end{eqnarray}
where we have written $k=2 \pi/\lambda = 2 \pi \nu n/c$. Recall that the `$\pm$' is positive when the acceleration is parallel to the motion (acceleration from rest), and negative when the acceleration is anti-parallel to the motion (acceleration to rest).

\subsubsection{Time-domain derivation}

For the time-domain derivation, we again consider the radiated component of Eq. \ref{lweqn1}. We can calculate the time-integral of the electric field for one starting point or stopping point, taking into account the conversion from retarded emission time $\tp$ to observer-time $t$ as $t = \tp + n R/c$ and $\dt = \dtp (1 - n \vec{\beta} \cdot \hat{r})$, via:
\begin{eqnarray}\label{time_derivation}
\int \vec{E}(\vec{x},t)\ \dt & = & \frac{q}{c} \int_{\Delta t} \left[ \frac{\hat{r} \times [(\hat{r} - n \vec{\beta}) \times \dot{\vec{\beta}}]}{(1 - n \vec{\beta} \cdot \hat{r})^3 R} \right]_{\rm ret} \dt \nonumber \\
& = & \frac{q}{c} \int_{\Delta \tp} \frac{\hat{r} \times [(\hat{r} - n \vec{\beta}) \times \dot{\vec{\beta}}]}{(1 - n \vec{\beta} \cdot \hat{r})^2 R}\ \dtp \nonumber \\
& = & \frac{q}{c} \int_{\tp_0}^{\tp_1} \frac{\mathrm{d}}{\dtp}\left( \frac{\hat{r} \times [\hat{r} \times \vec{\beta}]}{(1 - n \vec{\beta} \cdot \hat{r}) R} \right) \dtp \nonumber \\
& = & \pm \frac{q}{c} \left( \frac{\hat{r} \times [\hat{r} \times \vec{\beta}^*]}{(1 - n \vec{\beta}^* \cdot \hat{r}) R} \right) 
\end{eqnarray}
Here, $\Delta t = t_1-t_0$ denotes the observer-time window corresponding to the retarded-time window $\Delta \tp = t_1^{\prime}-t_{0}^{\prime}$, which encompasses the acceleration process. For a starting point ($+$ sign), the particle is at rest at the time $\tp_0$ and has velocity $\vec{\beta}^*$ at $\tp_1$. The opposite is the case for a stopping point ($-$ sign).

Since the acceleration is instantaneous, the distance $R_{\rm acc}$ from the particle to the observer at the acceleration time is constant, and the time $t_{\rm acc}$ at which an observer would view the radiation emitted at time $t^{\prime}_{\rm acc}$ is given by $t_{\rm acc} = t^{\prime}_{\rm acc} + n R_{\rm acc}/c$. The time window $\Delta t = t_1-t_0$ in Eq.\ \ref{time_derivation} is therefore chosen to satisfy $t_0 < t_{\rm acc} < t_1$.

While the electric field as a function of time $\vec{E}(\vec{x},t)$ becomes infinite in the case of instantaneous acceleration, the time-integrated electric field is finite and independent of the specific choice of $\Delta t$. Consequently, one can calculate the time-averaged electric field over the time-scale $\Delta t$ as
\begin{equation}\label{endpoint_eqn_t}
\vec{E}_{\pm}(\vec{x},t) = \pm \frac{1}{\Delta t}\frac{q}{c} \left( \frac{\hat{r} \times [\hat{r} \times \vec{\beta}^*]}{(1 - n \vec{\beta}^* \cdot \hat{r}) R} \right).
\end{equation}
An adequate choice of $\Delta t$ is dictated by the time resolution of interest. If $\Delta t$ is chosen significantly longer than the time-scale over which the acceleration process occurs --- which is in particular the case for the instantaneous acceleration considered here --- the details of the acceleration process are of no importance.

At first glance, the results given in Eqs.\ \ref{endpoint_eqn_f} and \ref{endpoint_eqn_t} for a radiating endpoint may appear as yet another special case of particle motion with very limited application. However, observe that in arriving at Eqs.\ \ref{endpoint_eqn_f} and \ref{endpoint_eqn_t}, we have made no assumptions about the macroscopic motion of the particle --- only that at a given instant, the particle becomes accelerated. As we will see, validating this assumption is really a question of describing the particle motion with sufficient accuracy for the frequency-range/time-resolution of interest, rather than being a limitation of the endpoint approach. In following sections, we will show how arbitrary particle motion can be described in terms of such endpoints. However, before proceeding to more complex situations, it is worthwhile examining the radiation from the most simple acceleration event, a single endpoint.

\subsubsection{Radiation pattern of a single endpoint}

The radiation pattern from a single endpoint is exactly that corresponding to a once-off acceleration event. A relevant physical situation would be the $\beta$-decay of a heavy element in vacuum, where the motion of the heavy nucleus can be neglected, and the emitted $e^{\pm}$ travels with constant velocity to infinity. There are quite a few interesting features of even this simple situation which are worthwhile to explore in greater depth.

For most applications, it is preferable to use the vectorial notation given in Eqs.\ \ref{endpoint_eqn_f} and \ref{endpoint_eqn_t} to describe the radiation from a single endpoint. However, for a single event, the radiation is cylindrically symmetric about the acceleration/velocity axis, so it is common to express these equations using an observer's position described by a distance $R$ and angle to the acceleration vector $\theta$ ($\theta=0 \Rightarrow \hat{r} \parallel \vec{\beta}^*$). This angular dependence is seen easily from the LHS of Fig.\ \ref{synch_diagram}. For this case, the magnitude of the electric field vector in Eqs.\ \ref{endpoint_eqn_f} and \ref{endpoint_eqn_t} respectively becomes:
\begin{eqnarray}
\vec{E}_{\pm}(\vec{x},\nu) & = & \pm \frac{q }{c} \, \frac{e^{i k R}}{R} \, \frac{\beta^* \sin \theta \ e^{2 \pi i \nu \tpo} }{1- n \beta^* \cos \theta} \, \hat{E}_{\pm} \label{sincos_endpoint_eqn_f} \\
\vec{E}_{\pm}(\vec{x},t) & = & \pm \frac{1}{\Delta t} \frac{q}{c} \, \frac{\beta^* \sin \theta}{(1- n \beta^* \cos \theta)R} \, \hat{E}_{\pm} \label{sincos_endpoint_eqn_t}
\end{eqnarray}
and it is taken as given that the unit electric field vector $\hat{E_\pm}$ points away from the acceleration axis for $\theta < \pi/2$ and towards it for $\theta > \pi/2$. At all times the angle $\theta$ is defined to be positive in the direction of positive velocity, irrespective of the acceleration. Thus under the transformation $\theta \to \pi-\theta$, $\beta \to - \beta$, Eqs.\ \ref{endpoint_eqn_f} and \ref{endpoint_eqn_t} are invariant, since $\hat{E}_{\pm} \to - \hat{E}_{\pm}$.

\begin{figure*}
\begin{center}
\includegraphics[width=0.49\textwidth]{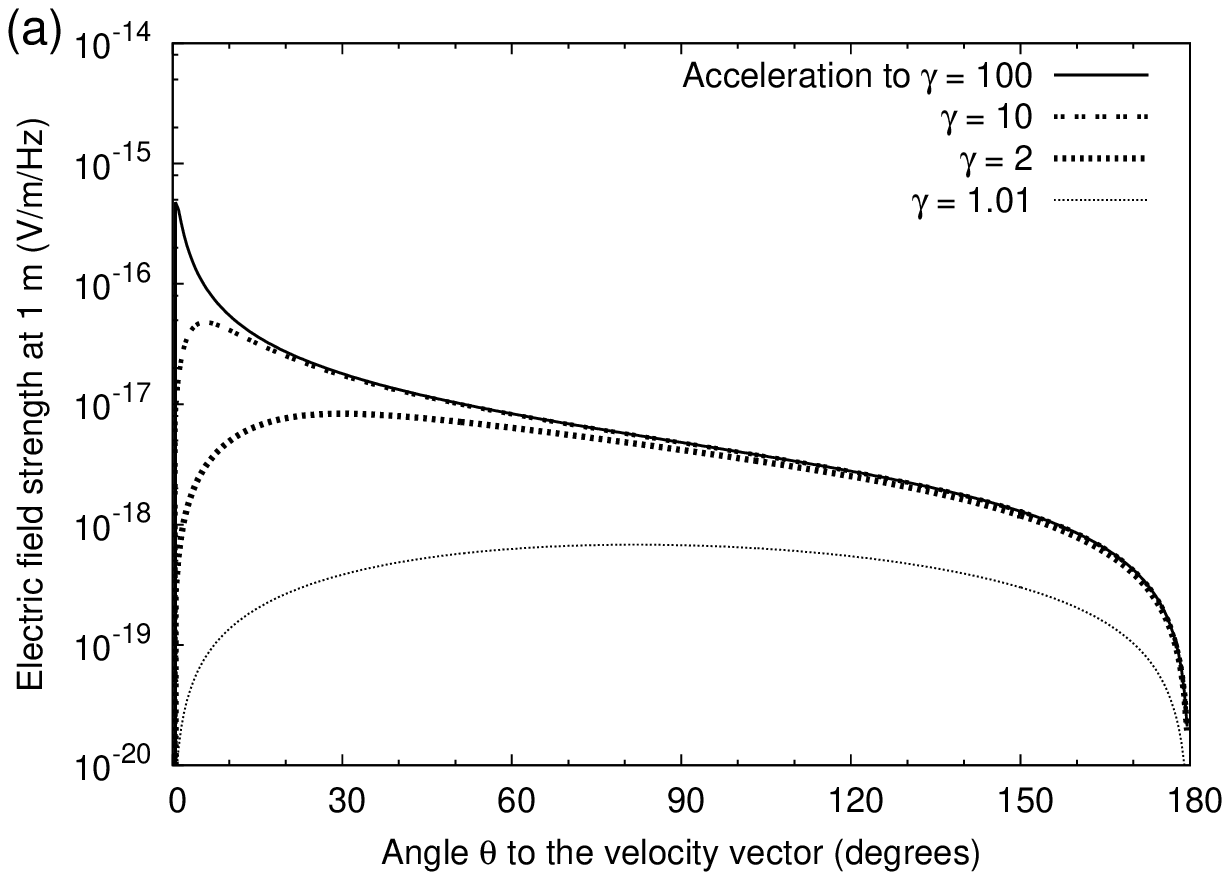} \includegraphics[width=0.49\textwidth]{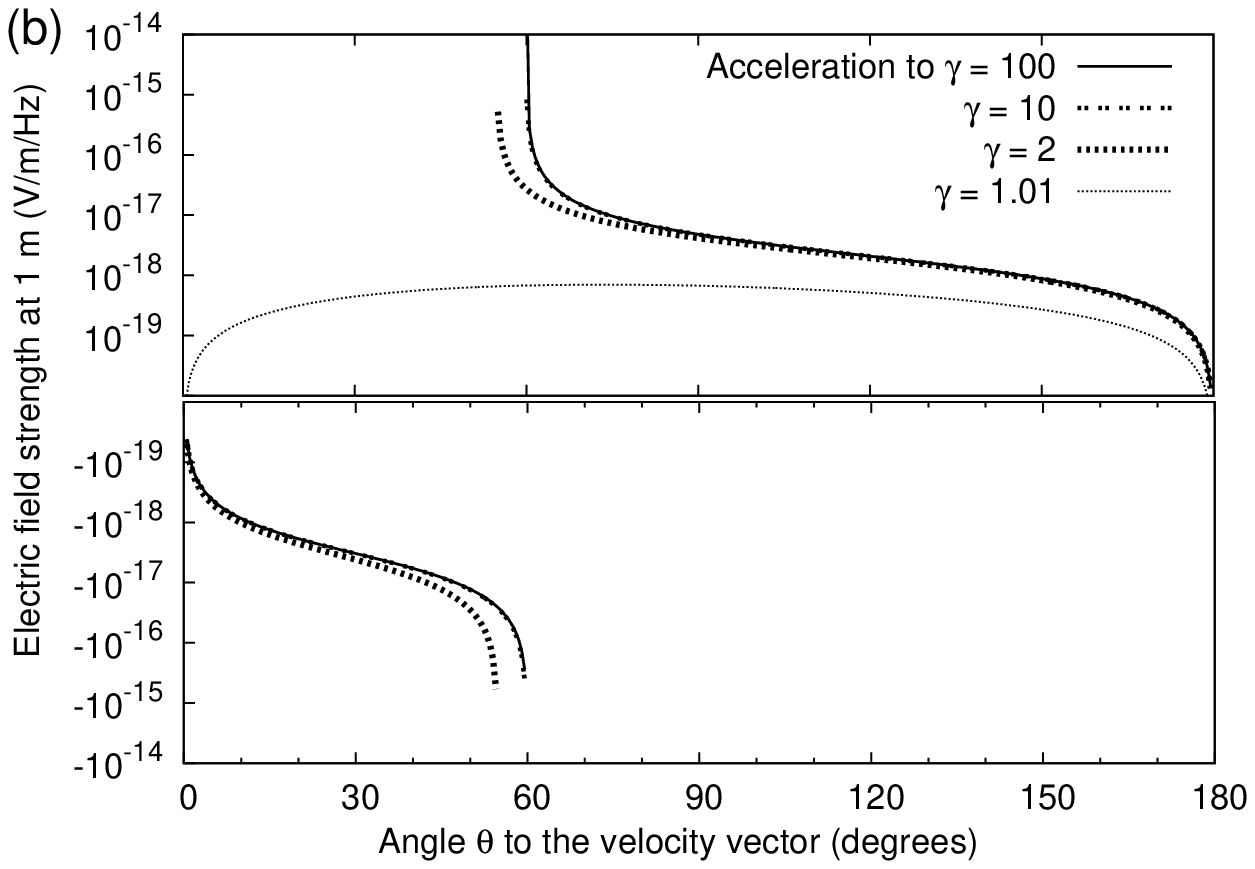}
\caption{Radiation from a single endpoint in vacuum as given by Eqs.\ \ref{sincos_endpoint_eqn_f}, and \ref{sincos_endpoint_eqn_t} multiplied by the time-interval $\Delta t$, as a function of the angle $\theta$ from the shower axis, for a range of $\gamma$-factors (and hence $\beta^*$s), left: in vacuum, and right: in an $n=2$ medium. In both figures, the highly relativistic cases produce near-identical radiation patterns.}
\label{single_endpoint_fig}
\end{center}
\end{figure*}

To illustrate, Eqs.\ \ref{sincos_endpoint_eqn_f}/\ref{sincos_endpoint_eqn_t} have been plotted in a vacuum and dielectric for varying $\beta$ in Fig.\ \ref{single_endpoint_fig}.

Firstly, note that for a single endpoint, the magnitude of the radiation in Eqs.\ \ref{sincos_endpoint_eqn_f}, \ref{sincos_endpoint_eqn_t} has {\it no} frequency-dependence. This may seem counter-intuitive, since almost all radiation processes become characterised by their particular frequency-dependence. Such frequency-dependence can only be produced however by the particle acceleration appearing differently on different wave-length scales, while a point-like acceleration looks identical on all scales, so that the resulting radiation could not possibly have any dependence on the wavelength/frequency. Only in the quantum-mechanical (extremely-high-frequency regime --- see Sec.\ \ref{discussion}) will there be a frequency-dependence in the radiation from a single endpoint, since the particle will no longer appear point-like.

Secondly, observe that there is a singularity in the emitted electric field about $n \beta^* \cos \theta = 1$ --- this is the `Cherenkov' singularity, which occurs at the Cherenkov angle $\theta_C = cos^{-1}(1/n \beta^*)$. Here, the electric field strength becomes undefined. This is, of course, unphysical, since we do not observe infinite electric fields in nature. Nonetheless, both Eqs.\ \ref{sincos_endpoint_eqn_f}, \ref{sincos_endpoint_eqn_t} and reality can happily coexist since an observer will always observe the particle traversing some finite observation angle $\delta \theta$. Writing $\theta = \theta_C + \delta \theta$, the divergent term in Eqs.\ \ref{sincos_endpoint_eqn_f} and \ref{sincos_endpoint_eqn_t} can be expanded in the vicinity of $\theta_C$ as follows:
\begin{eqnarray}
\frac{\beta^* \sin \theta }{R (1 - n \beta^* \cos \theta)} \approx \frac{1}{R(\theta_C) n \delta \theta} + \frac{1.5}{R(\theta_C) n^2 \beta^* \sin \theta_C}
\end{eqnarray}
The first term on the RHS, which diverges as $\delta \theta \rightarrow 0$, is odd about $\delta \theta = 0$, while the second (even) term is finite. Therefore, for any real measurement, an integral of the field about $\theta_C$ will have the divergent component cancel, leaving a finite result. In addition, any real medium will have a frequency-dependent refractive index, so that infinite field strengths will only be observed over an infinitely small bandwidth.

Finally, note that away from the singularity, there is a broad angular dependence which depends primarily on $\sin \theta$ and $\beta^*$. There is no emission in the exact forward direction for any values of $\beta^*$ and $n$, though for highly-relativistic particles in vacuum, the radiation pattern rises extremely rapidly away from $\theta=0$, producing the characteristic forward `beaming' expected. Also note that for mildly- and sub-relativistic particles, the emitted radiation at all angles changes with the particle energy, whilst in the ultra-relativistic regime, only extremely near to $\theta_C$ does the radiation pattern change with energy.

\subsection{Building physical situations}

\label{building}

We have derived both vectorial (Eq.\ \ref{endpoint_eqn_f} and Eq.\ \ref{endpoint_eqn_t}, in terms of $\vec{\beta}^*$, $\hat{r}$, $n$, $t_0^{\prime}$, and $R$) and scalar (Eq.\ \ref{sincos_endpoint_eqn_f} and Eq.\ \ref{sincos_endpoint_eqn_t}, in terms of $\beta^*$ $\theta$, $n$, $t_0^{\prime}$, and $R$) equations for the radiation from an endpoint. For the sake of brevity, in this section we use only the scalar notation of Eq.\ \ref{sincos_endpoint_eqn_f} and Eq.\ \ref{sincos_endpoint_eqn_t} to describe the situation.

Despite Eq.\ \ref{sincos_endpoint_eqn_f} and Eq.\ \ref{sincos_endpoint_eqn_t} describing the radiation resulting from a particle accelerating from/to rest, they are, in fact, more general than this. This is because an arbitrary acceleration of a particle can be viewed as a superposition of deceleration and acceleration events, which will not cancel if either $\theta$, $\beta^*$, or $n$ differ between the two endpoints\footnote{The superposition of two endpoints in this manner produces the ``well-known formula for radiation of charges which change their velocity sharply'' as discussed by Ginzburg (1982) \cite{Ginzburg82} and Bolotovskii, Davydov, and Rok (1982) \cite{BDR82}.}.
For curved particle motion at constant speed, the angle $\theta$ from the acceleration vector to the observer will be different for coincident endpoints, while for gradually accelerating/decelerating particles, the values of $\beta^*$ will be different for simultaneous endpoints. In either case, contributions from starting and stopping points will not cancel, and radiation will occur. Conversely, if a simple linear motion with constant velocity is described piece-wise as a series of starting and stopping endpoints, the terms will cancel completely --- the particle will not radiate. Superposition of endpoints in this way is sometimes viewed as destroying the `old' particle and creating a `new' one --- since this formulation is applicable only to the radiated component, static fields (which fall as $1/R^2$) can be ignored, so that bringing a particle to rest (`stopping' it) is equivalent to destroying it, and accelerating a particle from rest (`starting' it) is equivalent to creating it, and {\it vice versa}. An arbitrary change in particle velocity can be dealt with by combining two simultaneous, coincident endpoints, the first to `stop' the particle by bringing it from its old velocity to rest, the second to `start' the particle by accelerating it to its new velocity. Multiple particles/events can be treated by adding the contributions with appropriate $\theta$, $\beta^*$, $n$, $R$, and $t_0^{\prime}$. In the case of a smoothly-varying velocity $\vec{\beta}$, the values $\vec{\beta}^*$ used at the endpoints should be representative of the average velocity between endpoints, which will tend towards the true value of $\vec{\beta}$ as the number of endpoints used becomes large. Any propagation effects between the source and the observer --- e.g.\ absorption in a medium, or transmission through an interface --- should be applied to the (spherically-diverging) radiation from each endpoint. This can be simply done, since the relevant parameters (e.g.\ angle of incidence for transmission) will be uniquely defined for each such endpoint. Note that for ray-tracing methods, the rays will be diverging, and transmission problems should be handled accordingly.

\begin{figure*}
\includegraphics[width=0.49\textwidth]{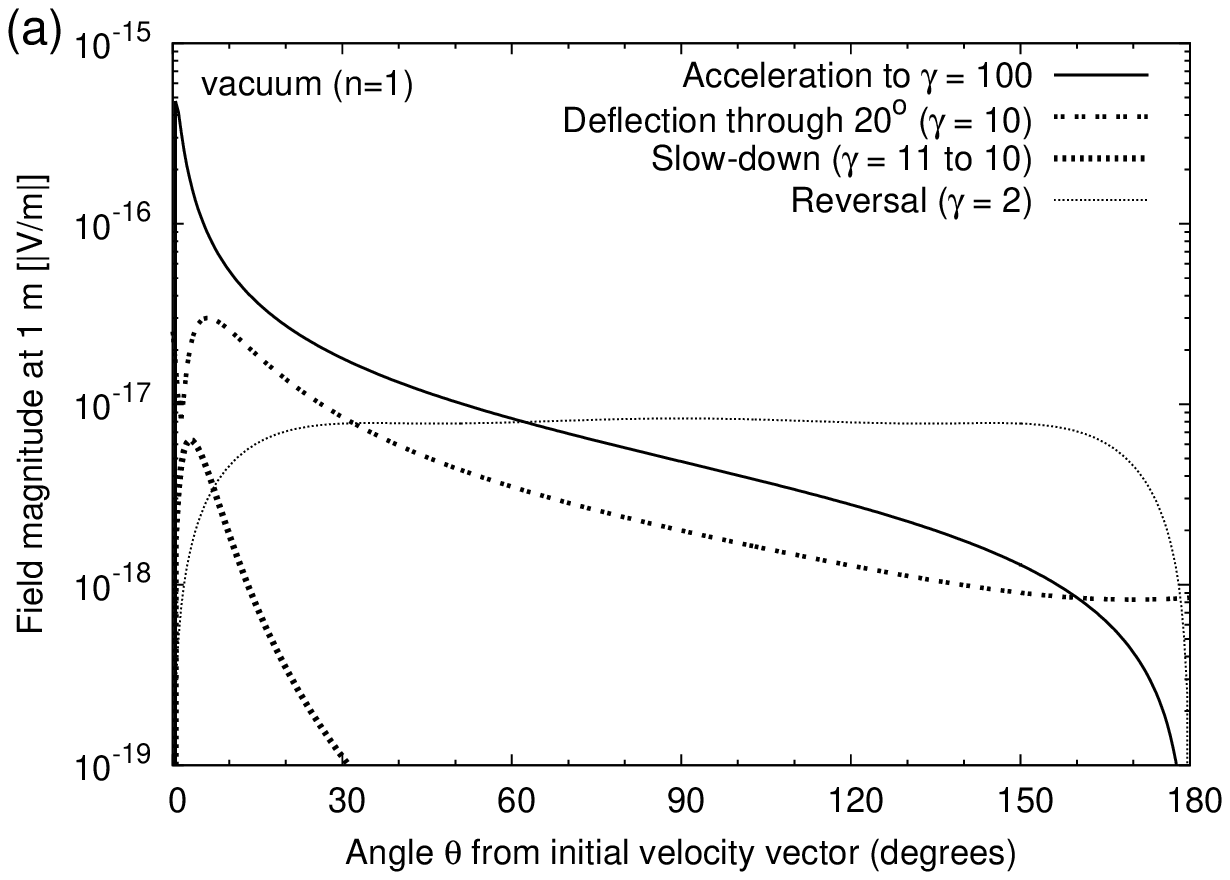} \includegraphics[width=0.49\textwidth]{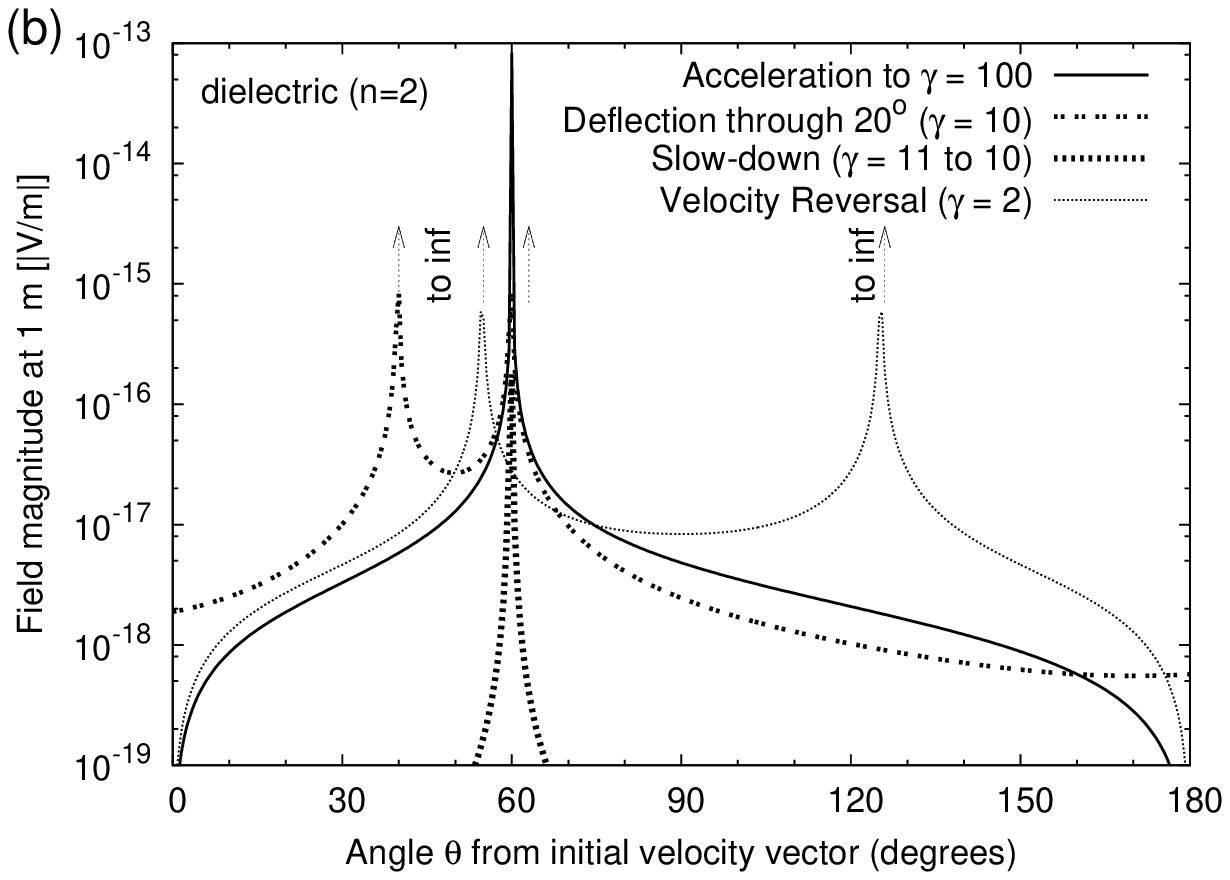}
\caption{Electric field magnitude resulting from the acceleration of a relativistic electron in four simple cases (see text) in (left) a vacuum and (right) a dielectric with refractive index $n=2$, as given by Eq.\ \ref{endpoint_eqn_f}.} \label{vac_di_fig}
\end{figure*}

To illustrate, we have plotted the emitted radiation in four elementary situations in a vacuum and dielectric in Fig.\ \ref{vac_di_fig}:  a single endpoint representing an electron accelerating from rest (`acceleration'); the deflection of an energetic electron through $20^{\circ}$ (`deflection'); the deceleration of a fast electron (`slow-down'); and a reversal of direction in a mildly relativistic electron with no change in speed (`reversal'). Note in the three highly relativistic cases the characteristic beaming --- in the forward direction for the vacuum case, and about the Cherenkov angle $\theta_C = cos^{-1} (1/(\beta^* n))$ in the $n=2$ dielectric. For velocity reversal, significant peaks are observed at the Cherenkov angle since $\beta^* n > 1$, while in the vacuum case, no appreciable beaming is evident and the emission is broad, which is characteristic of (non-relativistic) dipole radiation.

The simple examples presented in Fig.\ \ref{vac_di_fig} --- and Eqs.\ \ref{sincos_endpoint_eqn_f}, \ref{sincos_endpoint_eqn_t} themselves --- deal only with point-like acceleration events, while in most situations particle motion will be smooth; however, this is not a limitation in practice. Every numerical simulation necessarily describes particle motion as a series of uniform motions joined by instantaneous acceleration events, for which either of Eqs.\ \ref{endpoint_eqn_f} or \ref{endpoint_eqn_t} will calculate the emitted radiation {\it exactly}. The key point is that the degree to which the radiation calculated from the addition of endpoint contributions resembles the true radiation is limited only by the degree to which the simulated motion resembles the true motion. Usually this means that a particle simulator must be accurate to within a small fraction of the wavelengths of interest --- for a discussion of this effect in practice, see for example the discussion in Fig.\ 3 and Appendix A of Ref.\ \cite{AMVZ00} or section 3 of Ref.\ \cite{REAS3}. It is not a concern of this paper.

\subsubsection{Further notes on application}
\label{firstlastdiscussion}

In our experience of applying the endpoint formalism to complex physical situations, two important cases where the formalism can be mis-applied have come to our attention. We discuss each below.

The first case concerns the interpretation of the initial and final endpoints used to describe particle motion. If the initial endpoint is a starting/acceleration endpoint, this models the situation of a particle sitting at rest until suddenly accelerated, in which case the calculated radiation pattern will include a large contribution from this sudden acceleration. Therefore, this will be the correct choice if, in the physical situation being modelled, the particle genuinely does begin from rest. If it does not, the large initial contribution will be artificial and incorrect. If, on the other hand, the initial point is a stopping/deceleration endpoint, the implied motion is that of a particle moving with uniform velocity for an infinite time until the time of the initial endpoint. Given that such infinite uniform motion does not generally occur in nature, the usual interpretation for this choice is that of a particle beginning the calculation with a non-zero velocity, and that whatever motion it undertook before that point is not of interest to the calculation. The choice of final endpoint, obviously, has similar implications. For instance, in the case of synchrotron (curvature) radiation in Sec.\ \ref{synch_sec}, the initial point must be a stopping endpoint, and the final point a starting endpoint, since the radiation of interest is only that from the curved motion of the particle. However, for the radiation from a finite particle track in Sec.\ \ref{cherenkov}, the situation of interest really is that of a particle which accelerates from and decelerates to rest, and hence the initial and final endpoints are starting and stopping endpoints respectively.

The second case involves ensuring that the velocity $\vec{\beta}^*$ used at a starting endpoint and a subsequent stopping endpoint is consistent with the implied motion of the particle: if the particle is accelerated at a starting endpoint at $\vec{x}_1, t_1$ and arrives at a stopping endpoint at $\vec{x}_2, t_2$, then the condition $\vec{x}_{2} - \vec{x}_{1} = c\vec{\beta}^* (t_{2} - t_{1})$ must be satisfied. Furthermore, note that $\vec{\beta}^*$ must be identical at both the starting and stopping endpoints. If $\vec{\beta}^*$ changed between the starting and subsequent stopping point, this would imply an additional acceleration, and the radiation associated with this acceleration would not be accounted for, leading to incorrect results. This problem arises for an accelerating particle when the instantaneous particle positions and `true' velocities $\vec{\beta}$ are known (or simulated) at discrete times, so that in general the velocities $\vec{\beta}$ will not point towards the next known position. In such a case, one must realise that the velocities $\vec{\beta}^*$ used in the endpoint treatment are representative of the time-averaged true velocity $\vec{\beta}$ between endpoints. Therefore, the correct treatment is to re-normalise each instantaneous $\vec{\beta}$ (in both direction and magnitude) to the appropriate $\vec{\beta}^*$ to fulfill the above-mentioned condition. This will become important in the following section (Sec.\ \ref{synch_sec}) in the case of synchrotron radiation.

\section{Comparison to Established Theory}

\label{compare}

It is instructive to recreate classical radiating systems and reproduce the classical results using our endpoint formulation. We do this below for the cases of synchrotron, Vavilov-Cherenkov, and transition radiation.

\subsection{Synchrotron radiation}
\label{synch_sec}

\begin{figure}
\begin{center}
\includegraphics[width=0.15\textwidth]{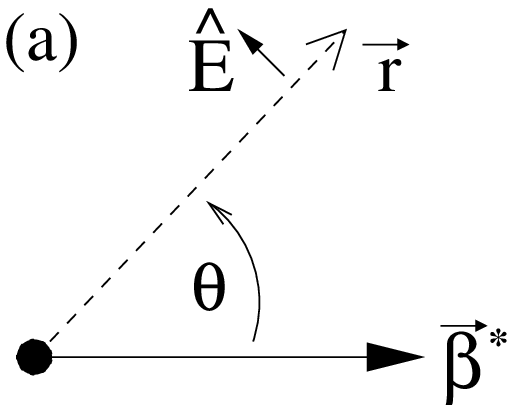} \vspace{0.2 cm} \includegraphics[width=0.3\textwidth]{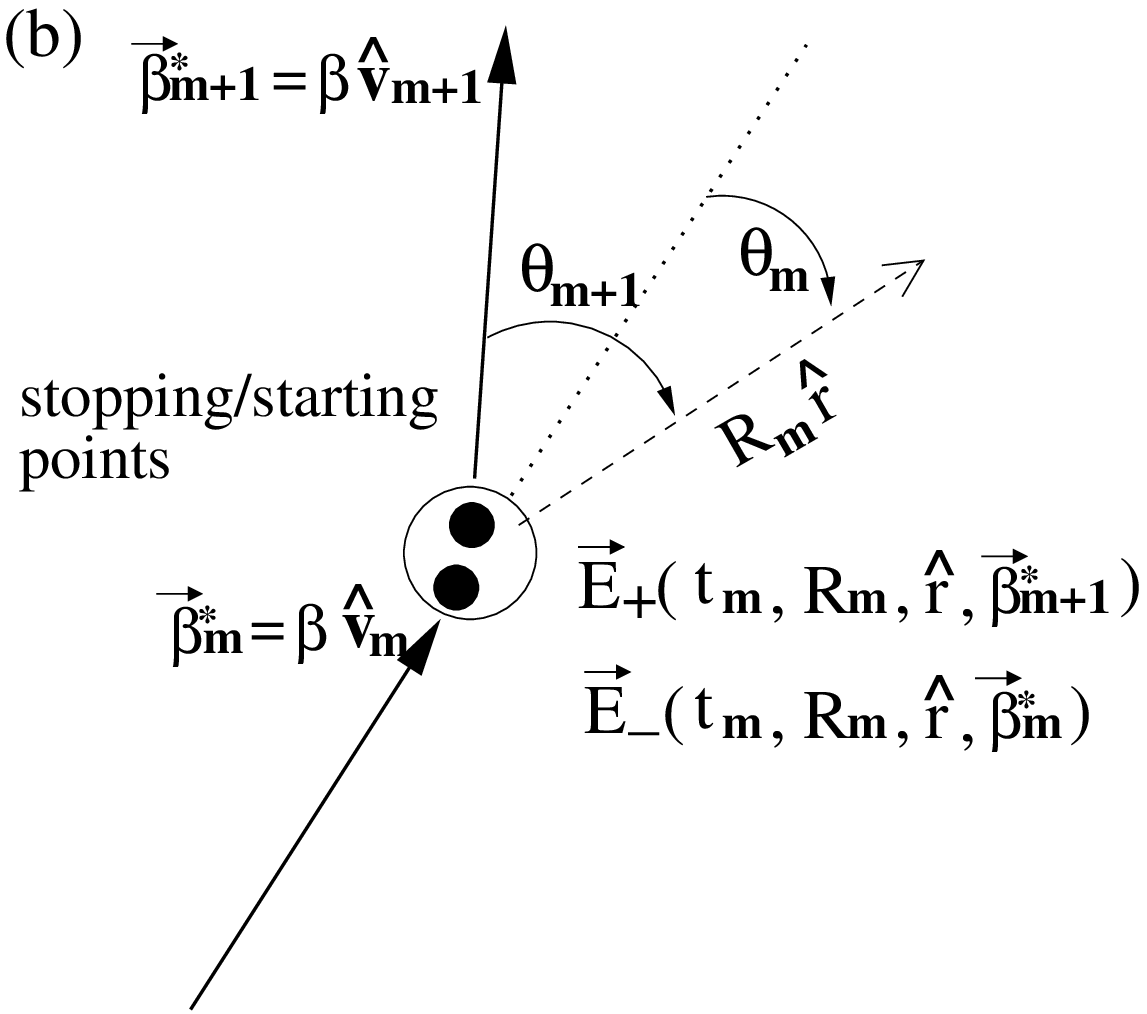}
\end{center}
\caption{(a) Sketch of $\theta$, the angle from the velocity vector; (b) Schematic diagram of the contributions from two terms in the sum of Eq.\ \ref{synch_eqn}. \label{synch_diagram}}
\end{figure}

Synchrotron radiation arises from a relativistic particle undergoing infinite helical motion (a superposition of circular motion in a 2D plane and linear motion perpendicular to the plane) in a vacuum ($n=1$), as is typically induced by the presence of a uniform magnetic field. Here, we treat the case of a particle of velocity $\beta$ executing a single circular loop of radius $L$ in the $x-y$ plane only in a vacuum. This motion is viewed by an observer at a very large distance $\sim R$ so that the unit-vector in the observer diretion $\hat{r}$ can be assumed to remain constant throughout the motion. Such motion can be represented by a series of $N$ starting and stopping points, schematically in Fig.\ \ref{synch_diagram}, and mathematically as follows:
\begin{eqnarray}
\sum_{m=0}^{N-1} \vec{E}_{+}(t_m, R_m, \hat{r}, \vec{\beta}^*_{m+1}) + \vec{E}_{-}(t_m, R_m, \hat{r}, \vec{\beta}^*_m), \label{synch_eqn}
\end{eqnarray}
where the calculation of each velocity vector $\vec{\beta}_m^*$, time $t_m$, and distance $R_m$ to the observer is a matter of simple geometry. While the $1/R_m$ term can be assumed constant, $R_m$ also changes the relative phase-factors between emission at different endpoints. Note that every starting term is balanced at any time $t_m$ by a simultaneous stopping term at the same position. The reason the terms do not cancel is due to the direction of the velocity vectors $\vec{\beta}^*_m = \beta^* \hat{v}_m$ differing between simultaneous starting and stopping terms. Note also that there is only one term with $\vec{\beta}^*_0$ (a stopping event) and one with $\vec{\beta}^*_N$ (a starting event). This represents the physical situation of a particle moving in some direction from $-\inf$, executing the loop described, then continuing on to $+\inf$ in the original direction, as described in more detail in Sec.\ \ref{firstlastdiscussion}. This is necessary so that the radiation modelled is due to the curvature of the particle (that is, synchrotron radiation), rather than any sudden and large initial and final accelerations to bring the particle from/to rest, which would tend to dominate. Finally, also note that the magnitude $\beta^*$ of the vectors $\vec{\beta}^*$ used in the calculation must be slightly decreased from the `true' value $\beta$, since the (straight line) distance between endpoints is slightly shorter than the distance along a circular arc; similarly, the vectors $\vec{\beta}^*$ will not be quite tangential. The necessity of this is also discussed in Sec.\ \ref{firstlastdiscussion}.

\begin{figure*}
\begin{center}
\includegraphics[width=0.485\textwidth]{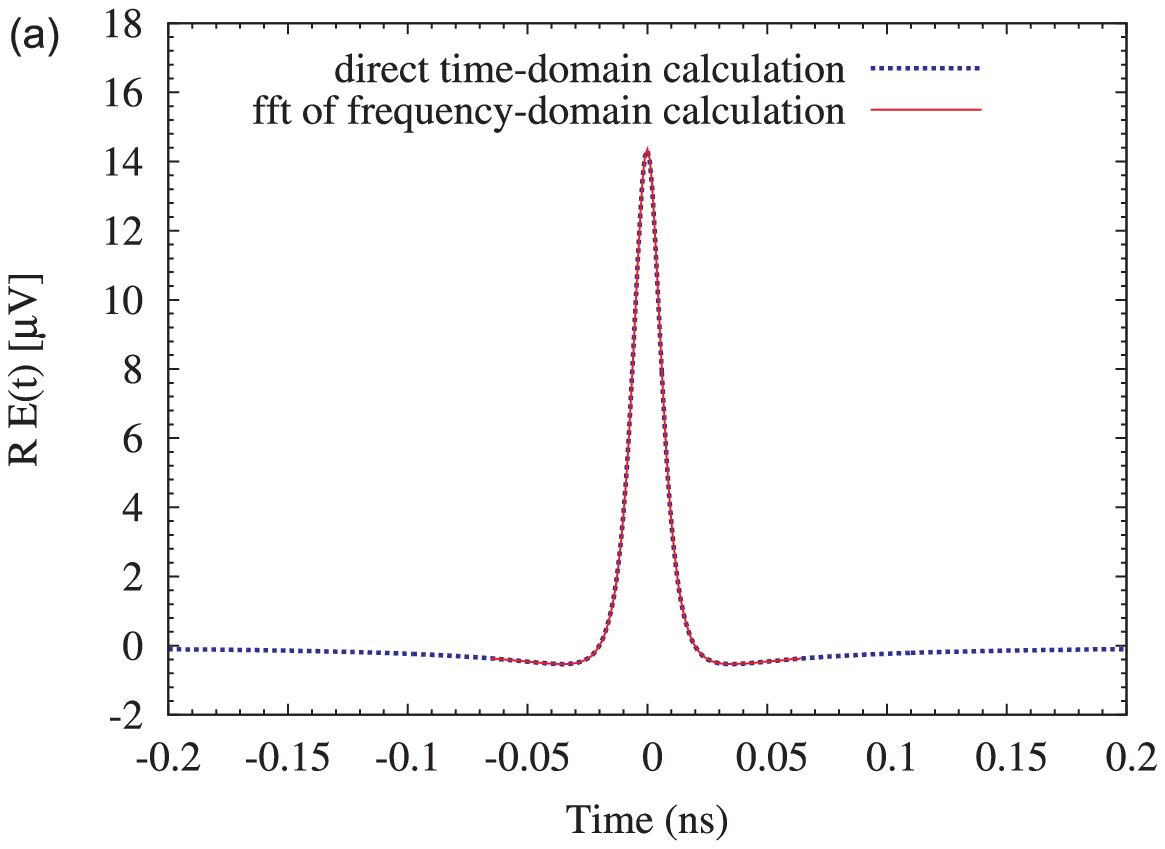} \includegraphics[width=0.50\textwidth]{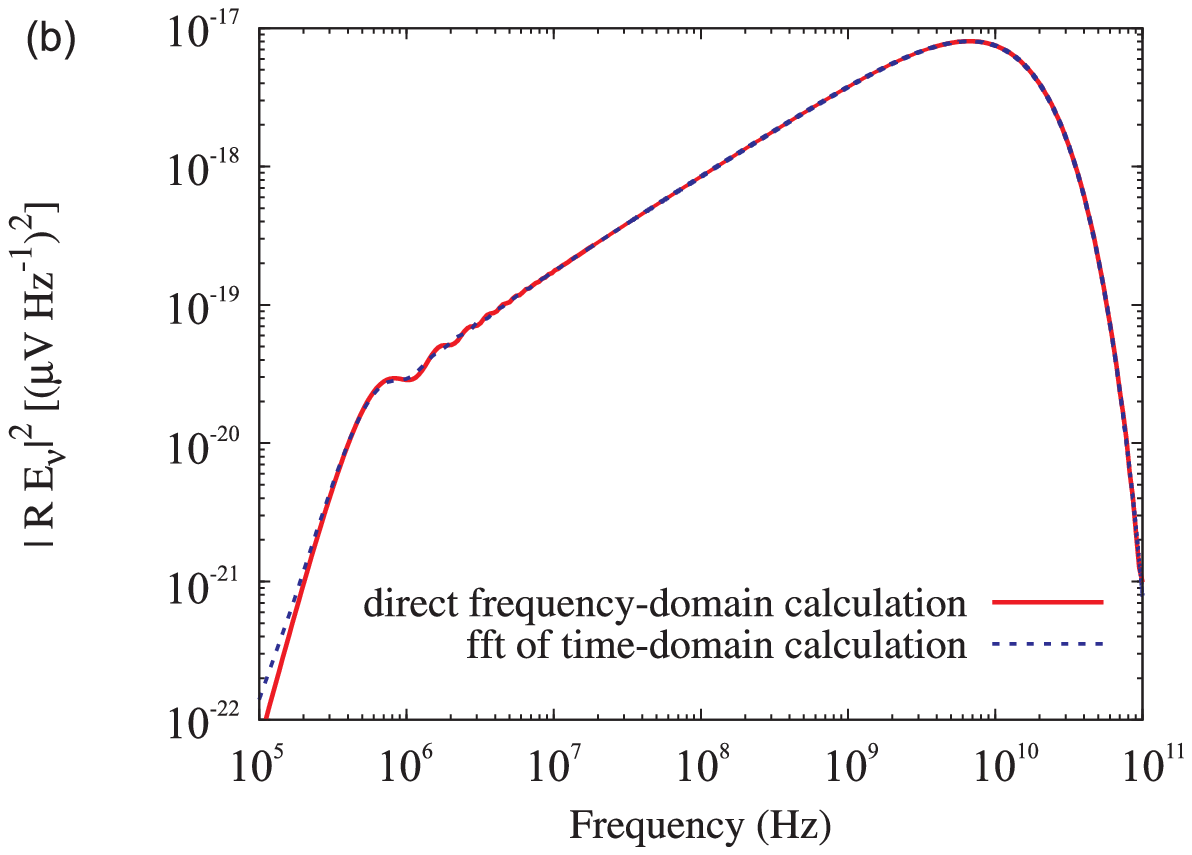}
\end{center}
\caption{(Color online) Time-trace of a synchrotron pulse produced by a single gyration of an electron with $\beta = 0.999$ and gyration radius $r=100$~m (left), and power spectrum of the same pulse (right). In each case, the direct calculations (time- and frequency-domain respectively) are shown in comparison with the results generated by fast-Fourier-transforming data from the other domain (frequency- and time-domain respectively). The direct calculation produces the higher-quality result and is much less computing-intensive. (For such practical reasons, the indirect, frequency-domain calculation of the time-trace was performed by modelling only the relevant part of the gyration cycle indicated by the plotted time range.) \label{synch_fig}}
\end{figure*}

We present numerical evaluations of Eq.\ \ref{synch_eqn} in Fig.\ \ref{synch_fig} for a loop of radius $L=100$~m and true velocity $\beta=0.999$ ($\gamma \approx 22.4$), equivalent to an $11.4$~MeV electron moving perpendicular to a $3.809$~Gauss field. The observer is assumed to lie in the very-far-field in the plane of the loop. For the plot in the time (frequency) domain, we present both a direct calculation, and the results from taking a Fourier transform from the calculation in the frequency (time) domain. All the characteristic features are reproduced perfectly: a steep spectral fall below the cyclotron frequency ($\nu = 2 \pi L (\beta c)^{-1}$), and a slow rise in power until an exponential cut-off above the critical frequency $\nu_{\rm crit} = 1.5 \gamma^3 \beta c/L \approx 50$~GHz. In the time-domain, a sharp pulse of characteristic width $1/\nu_{\rm crit}$ is seen. That the results calculated by Fourier transform do not exactly match the direct calculations is due to the difficulty in generating sufficient data to make an accurate transform. However, the correspondence is obvious. From here on therefore, we deal only with calculations in the frequency-domain, and take it for granted that one can transmute time-trace data to spectral data and {\it vice versa} accurately and as needed.

Perhaps the most familiar analytic result on synchrotron radiation is the normalised, angular-integrated
power spectrum for ultra-relativistic particles. This makes a useful comparison with our code and, by extension, the
endpoint formalism. The common analytic result writes the power spectrum  $P(\nu)$ as the product of a normalisation constant $C$ and
a dimensionless function $F(x)$, where $x=\nu/\nu_c$ is the ratio of the frequency $\nu$ to the critical synchrotron frequency
$\nu_c = 3 \beta c \gamma^3 / (2 \pi r)$ \cite{Jackson}:
\begin{eqnarray}
P(\nu) & = &  \frac{\sqrt{3} e^3 B}{m_e c^2} F(x), \\
F(x) & = &x \int_{x}^{\inf} K_{5/3}(\sqrt[3]{2.25 x^2}). \label{analytic_synch}
\end{eqnarray}
Here, $K$ is a modified Bessel function of the second kind. The 'ultra-relativistic approximation'
of $\beta = 1 $ in deriving this result means that it is only applicable in the frequency regime
far above the cyclotron frequency of $\nu_0 = \beta c / (2 \pi r)$, i.e.\ $x \gg \gamma^{-3}$.

\begin{figure*}
\begin{center}
\includegraphics[width=0.485\textwidth]{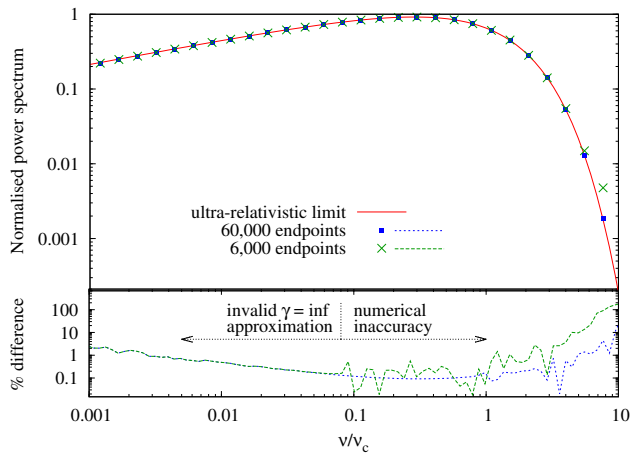}
\end{center}
\caption{(Color online) Top panel: normalised frequency spectrum from Fig.\ \ref{synch_fig}, calculated using $6 \times 10^4$ and $6 \times 10^3$ endpoints, and the analytic theory of the ultra-relativistic approximation. Bottom panel: the absolute relative difference between the ($\beta=0.999$) endpoint calculations and the ($\beta = 1$) theory. \label{synch_comparison_fig}}
\end{figure*}

To compare this analytic result with the endpoint theory calculation, we calculate the radiated far-fields as
above, but for all angles relative to the plane of gyration. These fields are then converted to
a radiated power and integrated over a $4 \pi$ solid angle, and plotted against $F(x)$
in Fig.\ \ref{synch_comparison_fig}\footnote{It is common in textbooks to plot a normalised $F(x)$ 
--- here, we do not include this normalisation factor, equal to $9 \sqrt{3}/(8 \pi)$.}.
Estimates using both $60,000$ and $6,000$
endpoints were made, to illustrate when numerical effects become important. The difference between the
analytic and endpoint method results is also shown, defined as
$2 |P_{\rm theory}-P_{endpoint}| / |P_{\rm theory}+P_{endpoint}|$. The difference in estimates at
high frequencies is due to numerical inaccuracy: this clearly reduces as the number of endpoints increases and better describes the curved track.
At low frequencies, the increasing disagreement with the analytic
result is due to the ultra-relativistic approximation breaking down, and here measures the
limitations in the theory, which will break down completely as
$x \rightarrow \gamma^{-3} \approx 8.9 \times 10^{-5}$.

\subsection{Particle tracks and Vavilov-Cherenkov radiation}

\label{cherenkov}

In this paper, we describe physical situations in terms of endpoints, whereas in numerical codes, the physical situation is usually described in terms of particle tracks. Such a track-based description defines the total charge distribution in position and time in terms of charges moving from position $(\vec{x}_1,t_1)$ to position $(\vec{x}_2, t_2)$. This implicitly defines a velocity $\vec{\beta}^*$; given the type of particle, the energy, momentum, and $\gamma$-factor are also defined. If the code outputs the positions `sufficiently' accurately, then the implied velocity $\vec{\beta}^*$ will also be `sufficiently' close to the true velocity $\vec{\beta}$ at each point.

It is obvious that the radiation from such a particle track can be constructed from two endpoint contributions --- one accelerating/creating a particle at $\vec{x}_1, t_1$ from rest to velocity $\vec{\beta}=\vec{\beta}^*$, the other decelerating/destroying the particle at $\vec{x}_2, t_2$. In the very-far-field of both events, the corresponding parameters $R, \theta$ can be written $\theta_1 = \theta$, $R_1=R$, $\theta_2=\theta$, $R_2 = R - \cos \theta \delta l$; we also write $t_2 = t_1+\delta t$. The track-length $\delta l$ is often expressed in terms of the time interval $\delta t$ via $\delta l = c \beta^* \delta t$. The entire particle track is considered as being sufficiently far from the observer so that the approximation $1/R_1 \approx 1/R_2 = 1/R$ holds \footnote{As discussed in detail by Afanasiev, Kartavenko, and Stepanovsky (1999) \cite{afanasiev99}, this approximation breaks down as the Cherenkov angle $\theta_C = \cos^{-1} (1/\beta^* n)$ is approached, since second-order terms in $R$ and $\cos \theta$ become important. It is exactly this approximation that removed the Vavilov-Cherenkov radiation component from Tamm's formula \cite{Tamm39} for the radiation from a finite particle track viewed at large distances, leaving only the `bremsstrahlung' contributions from the endpoints. However, this approximation --- and hence equation \ref{particle_track} --- is still valid for angles $\theta$ satisfying $|\theta -\theta_C| \gg \sin \theta_C \delta \ell R^{-1}$, where there is essentially no Vavilov-Cherenkov component.}. Again, it is more common to begin with the expression in Eq.\ \ref{sincos_endpoint_eqn_f}. Recalling $k = 2 \pi \nu n /c$, we find the radiation from a particle track to be:
\begin{eqnarray}
\vec{E}_{\rm track}(\vec{x}, \nu) & = & \frac{q}{ c} \frac{\beta^* \sin \theta}{1 - n \beta^* \cos \theta} \nonumber \\*
&& \cdot \left( \frac{e^{i (k R_1 + 2 \pi \nu t_1)}}{R_1} - \frac{e^{i (k R_2 + 2 \pi \nu t_2)}}{R_2} \right) \hat{E} \nonumber \\
& \approx & \frac{q \beta^* \sin \theta }{c} \frac{e^{i (k R + 2 \pi \nu t_1)}}{R} \nonumber \\*
&& \cdot \left( \frac{ 1 - e^{2 \pi i \nu (1- n \beta^* \cos \theta) \delta t}}{1 - n \beta^* \cos \theta} \right) \hat{E} \label{particle_track}
\end{eqnarray}
Note that Eq.\ \ref{particle_track} is (to within a factor of $2$, due to a different definition of the Fourier transform) the `Vavilov-Cherenkov radiation formula' of Eq.\ 12 in Zas, Halzen, and Stanev \cite{ZHS92}, with $\mu_r=1$ and $q=-e$. That is, although the ZHS code is commonly understood to calculate `the Vavilov-Cherenkov radiation' from a cascade in a dense medium, what it actually calculates is simply `the radiation' due to particle acceleration. When the particles in the cascade are all travelling in the same direction in a medium with refractive index significantly different from $1$, the radiation just so happens to very closely resemble the classical notion of Vavilov-Cherenkov radiation. The Vavilov-Cherenkov condition is plainly evident by letting the phase term $(1 - n \beta^* \cos \theta) \rightarrow 0$ --- that is, the observation angle $\theta$ tends towards the Vavilov-Cherenkov angle $\theta_C$, where $\cos \theta_C = (n \beta^*)^{-1}$ --- in which case Eq.\ \ref{particle_track} reduces to:
\begin{equation}
\vec{E}_{\rm track}(\vec{x}, \nu) =  2 \pi i \nu \delta t \, c \beta^* \sin \theta \frac{e}{c^2} \, \frac{e^{i (k R + 2 \pi \nu t)}}{R} \, \hat{E} \label{cherenkov_eqn}
\end{equation}
The product $c \beta^* \delta t = \delta l \sin \theta$, so that the radiated intensity is proportional to the apparent tracklength. Thus to a far-field observer near the Cherenkov angle, the radiation seen is consistent with emission per unit tracklength, although our description makes it clear that this is not the case\footnote{It was first noted by Zrelov and Ru\v{z}i\v{c}ka (1989) that Tamm's (1939) approximate formula for radiation from a finite particle track results from contributions from acceleration at the endpoints.}.

We conclude our discussion on Vavilov-Cherenkov radiation by noting that `true' Vavilov-Cherenkov radiation, which is emitted in the absence of particle acceleration in a dielectric medium, can not be dealt with by this methodology. This comes by virtue of the fact we deal only with the `radiation' component of the \LW potentials, whereas contributions from unaccelerated motion must come from the `nearfield' term. Thus in the classical treatment of Vavilov-Cherenkov radiation by Frank and Tamm \cite{FrankTamm37} in which an infinite particle track is assumed, the near-field must also be infinite. Thus it should come as no surprise that a radiation-based far-field treatment can not explain this phenomenon.

\subsection{Transition radiation}

\label{tr_section}

\begin{figure}
\begin{center}
\includegraphics[width=0.45\textwidth]{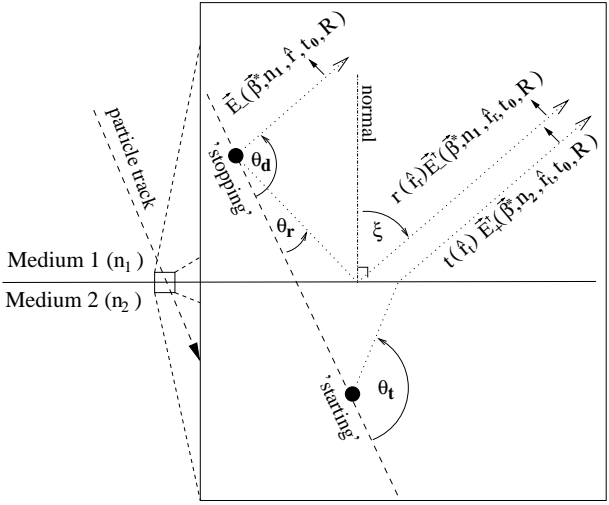}
\end{center}
\caption{Sketch of the endpoint formulation of transition radiation. See Eq.\ \ref{transition_eq} and the associated discussion. \label{tr_fig}}
\end{figure}

Transition radiation arises from a relativistic particle transitioning between two media with different refractive indices. The radiated energy was first calculated by Ginzburg and Frank \cite{GinzburgFrank44} in the case of a sharp boundary and a particle moving for an infinite time with uniform velocity parallel to the boundary surface normal.

J.~Bray (private communication) has noted that transition radiation can be explained as a particle being destroyed on the incoming side of the boundary, then being created on the outgoing side; this picture is consistent with Ginzburg's `mirror-charge' explanation of transition radiation, where the radiation in vacuum  from a particle entering an $n = \inf$ medium is explained by the sum of two charge contributions (real and mirror charges) which appear to disappear (or mutually annihilate) upon reaching the boundary \cite{Ginzburg82}. In terms of endpoint-theory, there exists a starting and a stopping event which are simultaneous and co-located (or rather, infinitesimally separated). The contributions from these events do not cancel because 1: the events occur in different media, and 2: an observer must be on one or the other side of the boundary, and thus the radiation from one of the two events will have to be transmitted through the boundary layer. The observed radiation will in fact be the sum of three contributions: a `direct' contribution from the event in the observer's medium, a `reflected' contribution off the boundary from the event in the observer's medium, and a `transmitted' contribution from the event in the non-observer medium. This situation is shown in Fig.\ \ref{tr_fig} for an observer in the `incoming' medium (medium $1$) with refractive index $n_1$ --- since the separation of the two endpoints is infinitely small, any such observer will be in the far field, and there will be no phase-offset between the three contributions. Using the endpoint formulation, the total field in medium $1$ is then:
\begin{eqnarray}
\vec{E}(\vec{x},t) & = & \vec{E}_{-} (\vec{\beta}^*, n_1, \hat{r}, t_0, R) \, + \, r(\hat{r}_r) \vec{E}_{-}^{\dag} (\vec{\beta}^*, n_1, \hat{r}_r, t_0, R) \nonumber \\
& & + ~ t(\hat{r}_t) \vec{E}_{+}^{\dag} (\vec{\beta}^*, n_2, \hat{r}_t, t_0, R). \label{transition_eq}
\end{eqnarray}
The first two terms arise from the particle in medium $1$ `stopping' at the boundary; while $\hat{r}$ is the usual direction towards the observer, $\hat{r}_r$ is the apparent observer direction seen in reflection from the boundary. The third term originates from the `starting' event in medium $2$, with $\hat{r}_t$ the apparent direction of the observer from the perspective of medium $2$. $r(\hat{r}_r)$ and $t(\hat{r}_t)$ are reflection and transmission coefficients, which vary according to the geometry defined by $\hat{r}_r$ and $\hat{r}_t$ and the relative refractive indices. In all terms, the velocity $\vec{\beta}^*$ should be that of the true velocity $\vec{\beta}$, since there is no possibility of acceleration between the two endpoints. The `$^{\dag}$' in the latter two terms is a reminder that field directions after reflection/transmission should be used. The field in medium $2$ can be expressed by switching $n_1 \leftrightarrow n_2$ and $\vec{E}_{-} \leftrightarrow \vec{E}_{+}$ in Eq.\ \ref{transition_eq}, and recalculating the vectors $\hat{r}, \hat{r}_r, \hat{r}_t$ and the transmission/reflection coefficients appropriately. This `three-contribution' formulation is the same as that noted by Ginzburg and Tsytovich \cite{GinzburgTsytovich} for the case of a particle normally incident at the boundary between two infinite uniform (but otherwise arbitrary) media.

In the case of normal incidence to a boundary, the geometry becomes substantially simpler. In this case, the observer direction is definable by the angle $\xi$ from the surface normal in the observer's medium (thus $0 \le \xi < \pi/2$), and all reflection/transmission will occur in the `parallel' direction. Note that $\xi$ is distinct from the angle $\theta$. Denoting the refractive index from the observer's medium as $n$, that from the other medium as $n^{\dag}$, and using $\pm$ ($\mp$) to indicate a `$+$' (`$-$') sign for observations in medium $1$ (the incoming medium) and a `$-$' (`$+$') sign for observations in medium $2$ (the outgoing medium), the scalar component of Eq.\ \ref{transition_eq} can be written thus:
\begin{eqnarray}
&& |\vec{E}(\xi, n, n^{\dag}, \beta^*)| = \frac{q}{c} \frac{1}{R(t_0^{\prime})} \label{tr_ep_eq2} \left(\mp \frac{\beta^* \sin \xi}{1 \pm n \beta^* \cos \xi} \right. \label{expanded_ep_tr_eq} \\
&& \mp \frac{r_{\parallel}(\xi,n ,n^{\dag}) \beta^* \sin \xi}{1 \mp n \beta^* \cos \xi}
\left. \pm \frac{ t_{\parallel}(\xi,n ,n^{\dag}) \beta^* \frac{n}{n^{\dag}} \sin \xi}{1 \pm n^{\dag} \beta^* \sqrt{1 - (\frac{n}{n^{\dag}} \sin \xi)^2}} \right) \nonumber
\end{eqnarray}
where
\begin{eqnarray}
r_{\parallel}(\xi, n ,n^{\dag}) & = & \frac{\cos \xi - \frac{n}{n^{\dag}} \sqrt{1-(\frac{n}{n^{\dag}} \sin \xi)^2}}{\cos \xi + \frac{n}{n^{\dag}} \sqrt{1-(\frac{n}{n^{\dag}} \sin \xi)^2}} \nonumber \\
t_{\parallel}(\xi, n ,n^{\dag}) & = & \frac{2 n \cos \xi}{n \sqrt{1 - (\frac{n}{n^{\dag}} \sin \xi)^2} + n^{\dag} \cos \xi} \label{coef_eq}
\end{eqnarray}
Observe that the reflection and transmission coefficients are those appropriate for a point source --- in the case of $r$, this is the same as that for a plane wave, while for transmission, the plane-wave result gets multiplied by $(n/n^{\dag}) \cos \xi (1 - (n/n^{\dag}) \sin^2 \xi)^{-0.5}$ (this can be thought of as accounting for the change in divergence of rays upon transmission). Note that the distance $R$ to the observer is constant for all contributions, since the two endpoints are infinitely close together, and that there is no explicit relative phase since the event times are also simultaneous (implicit phases can and do arise however, as we will see below).

For this situation, the radiated spectral energy density predicted by the endpoint formalism becomes:
\begin{eqnarray}
W(\nu, \xi) & = & 2 n \frac{c}{4 \pi} \left|R \, E(\xi, n, n^{\dag}, \beta^*)\right|^2 \label{sed_eq}
\end{eqnarray}
where the fore-factor of $2$ accounts for the energy radiated at negative frequencies. The frequency-dependence of the RHS of Eq.\ \ref{sed_eq} is contained in the implicit frequency-dependence of $n$ and $n^{\dag}$.

For the same physical situation, Ginzburg and Tsytovich define medium $1$ by its relative permittivity and permeability $\epsilon_1, \mu_1$, and medium $2$ via $\epsilon_2, \mu_2$. Using the same notation for $\pm$ and $\mp$ as for Eq.\ \ref{tr_ep_eq2}, and using for $\epsilon$, $\mu$ the values in the observer's medium and for $\epsilon^{\dag}$, $\mu^{\dag}$ the values for the other (non-observer) medium, these authors derive the radiated spectral energy distribution $W(\nu, \xi)$ ($\equiv 2 \pi W(\omega, \xi)$ for angular frequency $\omega$) as:
\begin{eqnarray}
&& W(\nu, \xi) = \label{gb_tr_eq} \\
&& 2 \pi \frac{q^2 v^2 \mu \sqrt{\epsilon \mu} \, |\epsilon-\epsilon^{\dag}|^2 \sin^2 \xi \, \cos^2\xi}{\pi^2 c^3 \left| \epsilon^{\dag} \cos \xi + \sqrt{\epsilon/\mu} \sqrt{\epsilon^{\dag} \mu^{\dag} - \epsilon \mu \sin^2\xi }\right|^2} \nonumber \\
&& \cdot \frac{\left| 1 \pm \frac{v}{c} \sqrt{\epsilon^{\dag} \mu^{\dag} - \epsilon \mu \sin^2 \xi} - \frac{v^2}{c^2} \frac{\epsilon \mu - \epsilon^{\dag} \mu^{\dag}}{\epsilon - \epsilon^{\dag}} \epsilon \right|^2}{\left|[1 - \frac{v^2}{c^2} \, \epsilon \mu \cos^2\xi] [1 \pm \frac{v}{c} \sqrt{\epsilon^{\dag} \mu^{\dag} - \epsilon \mu \sin^2 \xi}]\right|^2}. \nonumber
\end{eqnarray}
An alternative expression --- Eq.\ 2.45e of Ref.\ \cite{GinzburgTsytovich} --- is also given by Ginzburg and Tsytovich, which is suggestively close in form to our Eq.\ \ref{tr_ep_eq2}\footnote{Ginzburg and Tsytovich go further and point out that their expression can also be derived ``using the expression for a jump in the Lienard-Vichert [sic] potentials'', thanking V.\ V.\ Kocharovskii and Vl.\ V.\ Kocharovskii for this point. Since one or more endpoints imply a sudden jump in the potentials, it is likely that the two derivations are very similar.}.

\begin{figure}
\begin{center}
\includegraphics[width=0.45\textwidth]{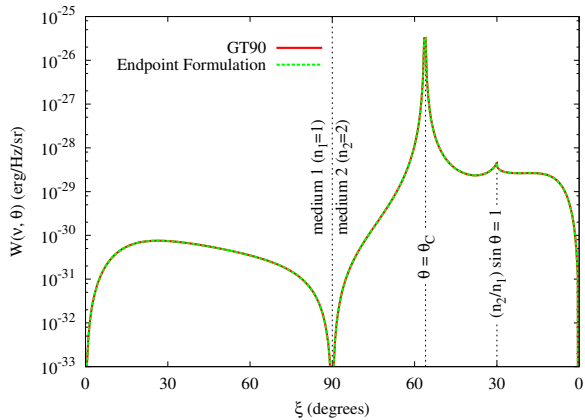}
\end{center}
\caption{(Color online) Radiated energy from the transition of a single electron/positron at normal incidence to the boundary between infinite dielectric media, calculated as per our `Endpoint Formulation' and using the result `GT90' of Ginzburg and Tsytovich for the spectral energy density. Only one line is visible since the results are in complete agreement.} \label{sed_fig}
\end{figure}

To compare the result given by the endpoint formulation (Eqs.\ \ref{expanded_ep_tr_eq}-\ref{sed_eq}) with that of Ginzburg and Tsytovich's Eq.\ \ref{gb_tr_eq}, we plot both results in Fig.\ \ref{sed_fig} for the case $\{\epsilon_1, \mu_1, \epsilon_2, \mu_2\} = \{1,1,4,1\}$ (i.e.\ $n_1=1$, $n_2 = 2$).

\begin{figure}
\begin{center}
\includegraphics[width=0.45\textwidth]{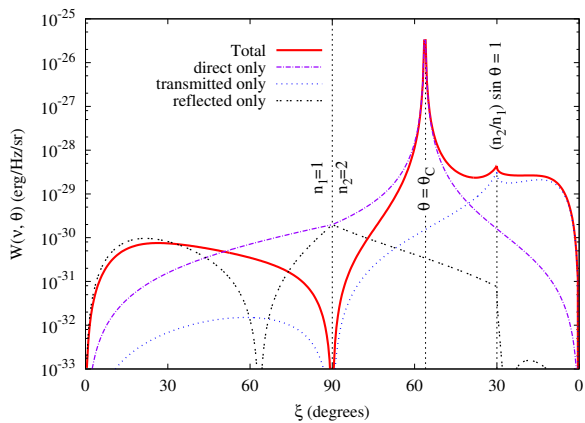}
\end{center}
\caption{(Color online) The radiated energy which would result from each of the three contributions --- direct, reflected, and transmitted --- if considered in isolation, compared to the total radiated energy, where the three field contributions are coherently summed before squaring.} \label{component_fig}
\end{figure}

Analysing Fig.\ \ref{sed_fig}, we observe that the loci overlap completely --- the endpoint formulation produces exactly the same result as Ginzburg and Tsytovich's (nine-page) derivation. The detailed angular-dependence of the spectrum arises from the angular-dependence of the three individual endpoint contributions, the behaviour of the reflection and transmission coefficients, and the interference between terms, which can be constructive or destructive. To illustrate, we plot in Fig.\ \ref{component_fig} the spectral energy density which would result from considering only one of the three terms --- direct, transmitted, or reflected --- in Eq.\ \ref{transition_eq}, as would be calculated by substituting the field contribution from that term only directly into Eq.\ \ref{sed_eq}. Regions of both constructive and destructive interference appear, and there are regimes in which each term dominates. In particular, note that it is the direct contribution which causes the large peak about the Cherenkov angle in medium $2$ (the Cherenkov condition is not met in medium $1$). This is another example where a peak at the Cherenkov angle does not imply the existence of `true' Vavilov-Cherenkov radiation (see discussion in Sec.\ \ref{cherenkov}). Interestingly, while the endpoint from which each contribution is derived changes across the boundary (e.g.\ the direct contribution comes from the stopping endpoint in medium $1$, and from the starting endpoint in medium $2$), the magnitude of each component is in fact continuous across the boundary.

A complication which bears discussing is that in the regime $90^{\circ} > \xi > 30^{\circ}$ of medium $2$, the `incident angle' of the transmitted component cannot be defined, since all incident angles from medium $1$ map to the range $30^{\circ} \ge \xi \ge 0^{\circ}$ in medium $2$. This is why, in Eqs.\ \ref{tr_ep_eq2} and \ref{coef_eq}, we have not used Snell's Law to define some $\sin \xi_i = (n/n^{\dag}) \sin \xi$ nor $\cos \xi_i = (1-(\frac{n}{n^{\dag}} \sin \xi)^2)^{0.5}$: in the regime $90^{\circ} > \xi > 30^{\circ}$ in medium $2$, $\sin \xi_i$ would be greater than one, and $\cos \xi_i$ purely imaginary, which does not make intuitive sense. The resolution to this dilemma is that the reflection and transmission coefficients result from solving continuity equations for the fields across a boundary. Solutions to these equations exist even when the solution is not that for incoming and outgoing plane waves in which Snell's law is usually defined. It is important in the endpoint formulation therefore to allow both reflection and transmission coefficients, and the contributions from individual endpoints, to be complex numbers. This is especially true when the refractive indices themselves can not be treated as purely real, as is the case for many applications of transition radiation on metallic targets. Finally, note that any and all frequency-dependence must come from the frequency-dependence of the refractive indices.

\section{Discussion}
\label{discussion}

The endpoint formalism described above provides a simple, accurate, and intuitive method for calculating the radiation resulting from particle acceleration. Using it, the radiated electromagnetic fields due to particle acceleration can be calculated in either the time- or frequency-domain for arbitrary particle motion. The domain in which to perform an endpoint-calculation should be the same as that of the desired result. While the process of fast-Fourier transforming between time- and frequency-domains is usually relatively quick, such transforms require an adequate number of points in the first domain to produce an accurate result in the second. This is especially true when a signal is localised in one domain (e.g.\ a short time-domain pulse, or a narrow-bandwidth signal), since then it will necessarily be spread over a great range in the other. Usually, if both the time- and frequency-domains are of interest, it will be computationally quicker to perform two direct calculations than to generate excess data points in one domain and use a fast-Fourier transform to convert to the other. Such was the case for the example of synchrotron radiation presented in Sec.\ \ref{synch_sec}. The only exception to this rule is when dispersive effects (changing refractive index with frequency) become important, in which case frequency-domain calculations would be more practical.

Like any method using a distribution of sources, the accuracy of the endpoint method will reflect the accuracy with which the distribution of endpoints reflects the true particle motion on scales of the smallest wavelength / highest time resolution of interest. With reasonable awareness of these issues however, our endpoint methodology can be used to calculate the radiation in some very complex physical situations, such as those described in Sec.\ \ref{applications}.

In emphasising the utility of the endpoint formulation, we should also mention its limitations, the most obvious of which is its classical foundation in Maxwell's equations: it breaks down in any quantum-mechanical limit. Specifically, it can not treat radiation processes involving only a single photon, nor the radiation of extremely energetic photons where the wavelength is of the order of the de Broglie wavelength of the radiating particle(s). Such limitations however are common to all classical methods of treating radiation and are not increased by our approach. The second limitation is that we have ignored the `nearfield' term from Eq.\ \ref{lweqn1}. This does {\it not} mean that our formulation cannot calculate radiation in the near-field of a source distribution. Since each endpoint is point-like, any observer is always in the far-field of any particular endpoint. Thus a near-field calculation requires only taking the trouble to re-calculate the direction to the observer from each endpoint individually. Only in certain special circumstances, such as the case of Vavilov-Cherenkov radiation from non-accelerated systems as discussed in Sec.\ \ref{cherenkov}, will the near-field term provide a significant contribution to the observed electric fields. In general, this nearfield term will only become important when a large part of the charge distribution passes very close to the detectors, and for most experiments it will represent at most a minor correction only.

Our last note is to emphasise that this paper is by no means the first to use (explicitly or implicitly) an endpoint-like treatment to solve for various radiation processes. The best-known endpoint-based treatment is the Larmor formula for the power radiated by an accelerated charge, which is commonly used in derivations of other radiation processes (again, see Jackson \cite{Jackson}). Also, as discussed in the introduction, there are numerous examples in the literature where multiple classical radiation processes have been described using the same fundamental underlying physics. What we have done here is to explicitly state that {\it all} radiation from particle acceleration can be described in terms of a superposition of instantaneous accelerations (endpoints), and give a general methodology for applying this method to an arbitrary problem.

\section{Practical Applications}
\label{applications}

\subsection{Coherent radio emission from near-surface cascades in the lunar regolith}

Strong pulses of coherent radiation are expected from extremely high energy (shower energy $E_s \gtrsim 10^{17}$~eV) particle cascades in dense media. The mechanism for producing the radiation is the Askaryan effect, whereby a total negative charge excess arises from the entrainment of medium electrons through e.g.\ Compton scattering, and the loss of cascade positrons via annihilation in flight \cite{Askaryan}. The radiation from the excess electrons travelling super-luminally through the medium will be coherent at wavelengths larger than the physical size of the cascade. The emission is the basis of the `Lunar technique' \cite{Dagkesamanskii}, a detection method by which the radiation is observed from ground-based radio-telescopes. Several current experiments utilise the technique \cite{LUNASKA, NuMoon, RESUNA}, which has been proposed to detect both cosmic-ray and neutrino interactions.

The emission from the Askaryan effect is considered to be coherent Vavilov-Cherenkov radiation, since this is the mode of radiation upon which Askaryan placed greatest emphasis in his papers \cite{Askaryan}, and the emission occurs in the case of charged particles moving super-luminally in a dielectric. If indeed this is the case, this would then lead to a formation-zone suppression of the radiation from near-surface cascades, such as those produced by cosmic rays, which interact immediately upon hitting the Moon. The reasoning is as follows:
consider such a near-surface cascade, induced by a cosmic-ray interaction near the regolith (dielectric)-vacuum boundary. It has been both predicted \cite{Ginzburg49, Ulrich66} and observed (qualitatively by Danos \etal\ \cite{Danos53}, and as a coherent pulse from single electron bunches by Takahashi \etal\ \cite{Takahashi}) that charged particles moving in a vacuum near a dielectric boundary generate Vavilov-Cherenkov radiation in the dielectric. Analogously, as the distance between the cascade and the surface tends to zero, the radiation emitted into the vacuum will approach that from a cascade in the vacuum itself (the `formation-zone' effect, first considered in this context by Gorham \etal\ \cite{Gorham_formation}). If indeed the emission from the Askaryan effect is Vavilov-Cherenkov radiation, which is generated by the passage of a particle through a dielectric, then since vacuum is not a dielectric, the emitted power into the vacuum from cascades nearing the boundary must tend towards zero. Thus the Askaryan emission from cosmic rays will be highly suppressed if the emission is from Vavilov-Cherenkov radiation.

However, the emission from the Askaryan effect is not in general Vavilov-Cherenkov radiation, because it arises from finite particle tracks viewed at a large distance. The emission is therefore of the same character as that produced in the `Tamm problem' of calculating the `Vavilov-Cherenkov radiation' from a finite particle track in a dielectric medium when viewed at a large distance \cite{Tamm39}. However, it has been pointed out by Zrelov and Ru\v{z}i\v{c}ka \cite{ZrelovRuzicka} that Tamm's 1939 result for the radiation in such a problem originates from the acceleration/deceleration at the beginning and end of the track. Thus neither Tamm's approximate result, nor the majority of the radiation emitted, is truly Vavilov-Cherenkov radiation\footnote{While as early as 1939 Tamm's calculations \cite{Tamm39} indicated that radiation from finite particle tracks would be different to that from infinite tracks (the 1937 Frank and Tamm result \cite{FrankTamm37}), it is this latter result, which gives zero emission in a vacuum, which is better known in the authors' fields, and commonly accepted to be `the' Vavilov-Cherenkov result. When Tamm's 1939 calculations are recalled, they are also treated as giving `Vavilov-Cherenkov' radiation, even though it can be shown \cite{afanasiev99} that the final result ignores the Vavilov-Cherenkov component.}. Given that the radiation from the Askaryan effect arises from the coherent superposition of radiation from many finite particle tracks, the majority of detectable radiation produced by the Askaryan Effect (in a dense $\rho \sim 1$~g/cm$^3$ medium) itself is not coherent Vavilov-Cherenkov radiation at all, but rather coherent radiation from particle acceleration. Macroscopically, the coherent radiation from microscopic accelerations and decelerations is correctly viewed as coming from the time-variation of net charge. Note that a shock wave at the Cherenkov angle will still be observed (this is also seen in transition radiation, as previously discussed in Sec.\ \ref{tr_section}). While Askaryan also mentioned the possibility of coherent transition radiation and `bremsstrahlung' (radiation from particle acceleration) in his first (1961) paper, the majority of the paper refers to Vavilov-Cherenkov radiation and coherent radiation from a moving charge excess, rather than to an accelerated charge excess.

Thus the primary reason why the zero-emission argument is incorrect is that the radiation due to the Askaryan Effect does not resemble what is commonly considered to be `true' Vavilov-Cherenkov radiation as described in the Frank-Tamm picture \cite{FrankTamm37}, which presents a negligible contribution in the Askaryan effect and indeed will tend towards zero as a particle cascade develops closer to a vacuum-interface. Instead of the above arguments, the use of our endpoint formulation much more easily resolves the issue: charged particles are accelerated and decelerated, ergo, the system radiates, and the effect of the nearby surface from the point of view of a far-field observer is no worse or more profound than for any transmission problem.

\subsection{Radiation from extensive air showers}

A second example is the case of radio emission from extensive air showers. When an energetic primary particle interacts in the upper atmosphere, it produces a cascade of secondary particles which can reach ground level. Radiation at frequencies of a few tens to a few hundreds of MHz from the electron/positron component of these cascades has been both predicted \cite{Askaryan,KahnLerche66,Colgate67,Allan71} and observed \cite{Jelley,LOPES,CODALEMA,Auger_radio}, with up until recently only fair agreement between predictions and measurements. The emitted radiation is often understood in terms of one or more classical radiation mechanisms, and it can be unclear as to what extent the mechanisms are different ways of explaining the same phenomena, or are truly separate effects.

The transverse current model \cite{KahnLerche66} describes the effect of the magnetic field as causing a macroscopic flow of charge (a transverse current) due to the different drift directions of electrons and positrons. The time-variation of this current in the course of the air shower evolution produces radiation polarized in the direction denoted by the Lorentz force. A modern implementation of the transverse current model, the MGMR model \cite{MGMR}, complements the transverse current emission with additional radiation components, in particular the emission from a relativistically moving dipole and from a time-varying charge excess, the latter of which essentially corresponds to the Askaryan effect. Difficulties can, however, arise in the separation of these phenomenological ``mechanisms''. For example, a component similar to Vavilov-Cherenkov radiation would appear even in case of charge-neutral particle showers in the presence of a magnetic field, because the magnetic field would induce a sufficient spatial separation of the positive and negative charges for the electron and positron contributions not to cancel at the frequencies of interest \cite{EGSnrc}.

In contrast to such macroscopic descriptions, microscopic Monte Carlo models calculate the radio emission as a superposition of radiation from individual charges being deflected in the geomagnetic field. This approach, although originally inspired by the notion of `geosynchrotron' radiation \cite{FalckeGorham03,HuegeFalcke03}, does in fact not need the assumption of any specific emission mechanism. Monte Carlo codes for the calculation of radio emission from extensive air showers have been realized by different authors in various time-domain implementations \cite{Suprun,REAS1,ReAires,REAS2}. It recently turned out, however, that all of these implementations (and others) in fact treated the emission physics inconsistently, thereby neglecting radio emission produced by the variation of the number of charged particles during the course of the air shower evolution. With the endpoint formalism described here, a new and fully consistent implementation of a microscopic modelling approach has been realized in the REAS3 code \cite{REAS3}. The universality of the endpoint formalism ensures that the radio emission from the motion of the charged particles is predicted in all of its complexity. In case of the REAS3 code, this becomes evident since both the ``transverse current'' radiation polarized in the direction of the Lorentz force and the radially polarized ``charge excess'' emission is reproduced automatically and in good agreement with macroscopic calculations \cite{Huege_ARENA}. The importance of a consistent treatment taking into account also the radiation due to the variation of the number of charged particles within an air shower is obvious when comparing results obtained with REAS3 and results obtained by the former implementation in REAS2.59. 
\begin{figure*}
\begin{center}
\includegraphics[angle = 270, width=0.485\textwidth]{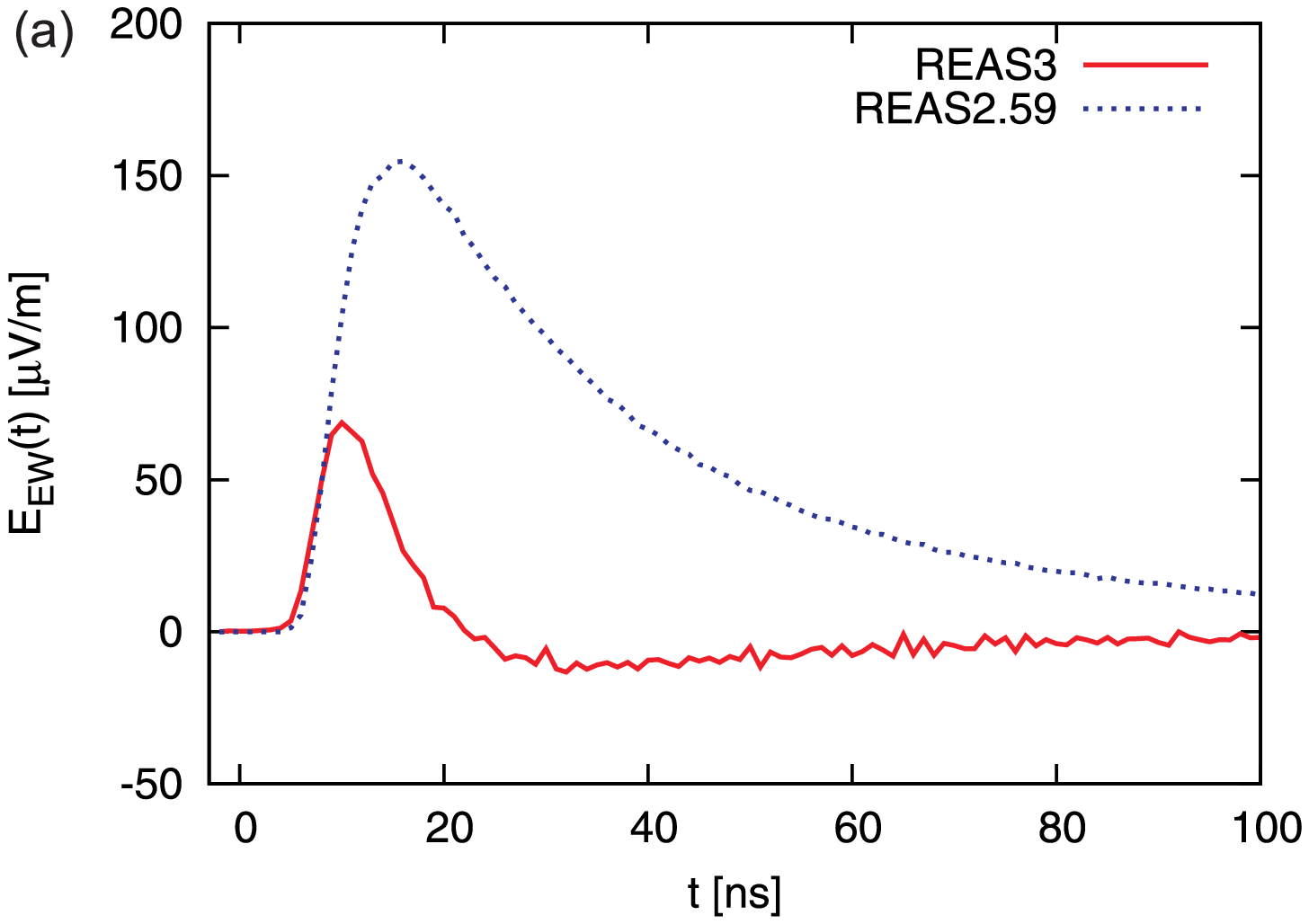} \includegraphics[angle = 270, width=0.485\textwidth]{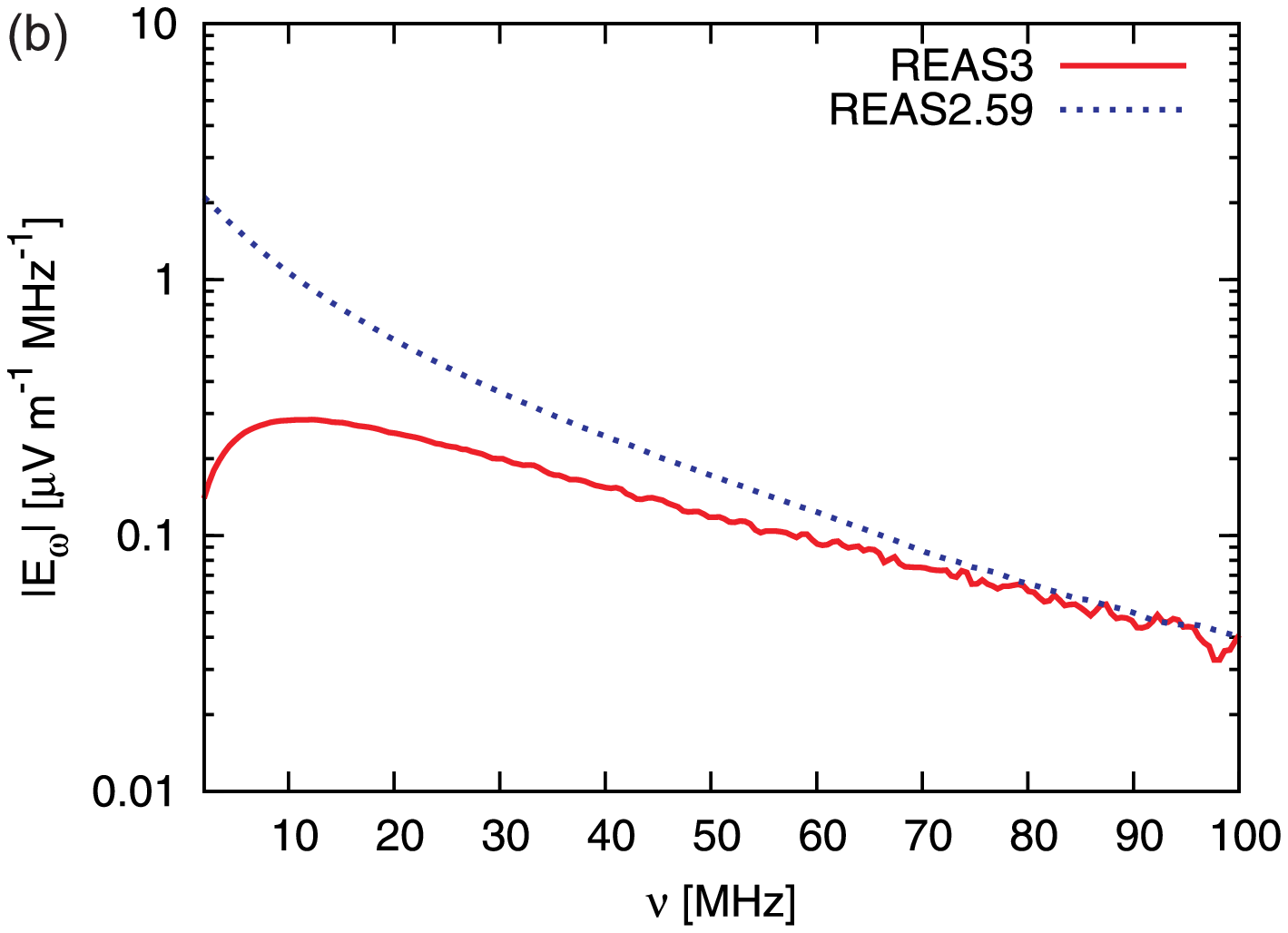}
\end{center}
\caption{(Color online) Simulations of the radio emission from a vertical extensive air shower with primary energy of $10^{17}$\,eV. The comparison of radio pulses (left) and corresponding frequency spectra (right) simulated with REAS3 and REAS2.59 at an observer position 200\,m south of the shower core illustrate the importance of a consistent treatment as implemented in REAS3 based on the endpoint formalism. The correctly predicted pulses become bipolar. Equivalently, the correct frequency spectra fall off to zero at zero frequency.\label{reas3_vs_reas2_pulse}}
\end{figure*}
This is illustrated for a specific example in Fig.\ \ref{reas3_vs_reas2_pulse} for the radio emission received by an observer 200\,m south of the core of a vertical extensive air shower with a primary particle energy of $10^{17}$\,eV. The pulse shapes, the pulse amplitudes and the frequency spectra differ significantly. For a detailed comparison of REAS3 and REAS2 we kindly refer the reader to \cite{REAS3}. The next step in improving the simulations will be the inclusion of the refractive index of the atmosphere, which is slightly different from unity and varies with atmospheric density.

\section{Conclusion}

We have presented  an `endpoint' methodology for modelling the electromagnetic radiation produced by the acceleration of charged particles. The approach is universally applicable and is especially well-suited for numerical implementation. Its universality has been illustrated by reproducing prototypical radiation processes such as synchrotron radiation, Tamm's description of Vavilov-Cherenkov radiation, and transition radiation in the frequency- and time-domains. The method's true strength, however, lies in modelling more complex (in other words ``realistic'') situations in which such individual prototypical radiation mechanisms can no longer be easily disentangled. As demonstrated in Sec.\ \ref{applications}, the `endpoint' methodology can for example be used to solve outstanding problems in the field of high-energy particle astrophysics. In conclusion, we would like to point out that we believe the `endpoint' approach to be an important way of viewing radiation processes which is useful at both an undergraduate student level, and also for career researchers. Except perhaps for those researchers working constantly with fundamental electromagnetic theory, we hope this methodology will increase the reader's understanding of radiative processes by providing a simple and unified approach. 

\acknowledgments{The authors would like to acknowledge J.~Alvarez-Mu\~{n}iz and J.~Bray for discussions regarding the `ZHS' code
and the nature of transition radiation. This research has been supported by grant number
VH-NG-413  of the Helmholtz Association. C.W.~James was the recipient of a 2009 Rubicon Fellowship from the NWO.}


\begin{thebibliography}{99}

\bibitem{Ginzburg82} V.L.~Ginzburg, Physica Scripta {\bf T2}, 182 (1982).

\bibitem{Sokolov77} A.A.~Sokolov, A.Kh.~Mussa, and Yu.G.~Pavlenko, Sov.Phys.J.\ {\bf 20}, 599 (1977). 

\bibitem{Pogorzelski74} R.J.~Pogorzelski, C.~Yeh, and K.F.~Casey, J.App.Phys.\ {\bf 45}, 5251 (1974).

\bibitem{Erteza62} A.~Erteza and J.J.~Newman, J.App.Phys.\ {\bf 33}, 1864 (1962). 

\bibitem{SchwingerTsaiErber} J.~Schwinger, W.~Tsai, and T.~Erber, Annals of Physics {\bf 96}, 303 (1976).

\bibitem{zhaires} J.~Alvarez-Mu\~{n}iz, W.R.~Carvalho, Jr, M. Tueros, and E. Zas,  arXiv:1005.0552 (2010).

\bibitem{Jackson} J.D.~Jackson, {\it Classical Electrodynamics} (John Wiley \& Sons, New York, 1998), $3^{\rm rd}$ ed.

\bibitem{AMRWZ2010} J.~Alvarez-Mu\~niz, A.~Romero-Wolf, E.~Zas, Phys.Rev.~D {\bf 81}, 123009 (2010).

\bibitem{BDR82} B.M.~Bolotovskii, V.A.~Davydov, and V.E.~Rok, Sov.Phys.Uspekhi {\bf 25} [3], 167 (1982).

\bibitem{AMVZ00} J.~Alvarez-Mu\~{n}iz, R.A.~V\'{a}zquez, and E.~Zas, Phys. Rev. D {\bf 62}, 063001 (2000).

\bibitem{REAS3} M.~Ludwig and T.~Huege, Astropart. Phys. {\bf 34}, 438 (2010)

\bibitem{afanasiev99} G.N.~Afanasiev, V.G.~Kartavenko, and Y.P.~Stepanovsky, J.Phys.D {\bf 32}, 2029 (1999).

\bibitem{Tamm39} I.E.~Tamm, J.Phys.\ (Moscow) {\bf 1}, 439 (1939).

\bibitem{ZHS92} E.~Zas, F.~Halzen, and T.~Stanev, Phys.~Rev.~D {\bf 45}, 362 (1992).

\bibitem{GinzburgFrank44} V.L.~Ginzburg and I.M.~Frank, Zh. Eksp. Teor. Fiz. [Sov. Phys. JETP] {\bf 16}, 15 (1946); I.M.~Frank and V.L.~Ginzburg, J.Phys.\ (Moscow) {\bf 9}, 353 (1945).

\bibitem{Askaryan} G.A.~Askaryan, Sov.~Phys.~JETP, \textbf{14}, 441 (1962); \textbf{48}, 988 (1965).

\bibitem{Dagkesamanskii} R.D.~Dagkesamanskii and I.~M.~Zheleznykh, Sov.Phys.\ JETP Let. {\bf 50}, 233 (1989).

\bibitem{LUNASKA} C.W.~James, R.D.~Ekers, J.~\'{A}lvarez-Mu\~{n}iz, J.D.~Bray, R.A.~McFadden, C.J.~Phillips, R.J.~Protheroe, and P.~Roberts, Phys.~Rev.~D {\bf 81}, 042003 (2010). 

\bibitem{NuMoon} O.~Scholten \etal, Phs.Rev.Lett.\ {\bf 103}, 191301 (2009). 

\bibitem{RESUNA} T.~Jaeger, R.L.~Mutel, and K.G.~Gayley, Astropart. Phys. {\bf 34}, 293 (2010).

\bibitem{Ginzburg49} V.L.~Ginzburg and I.M.~Frank, Dok.~Akad.~Nauk SSSR {\bf 56}, 699 (1947).

\bibitem{Ulrich66} R.~Ulrich, Zeitschrift f\:{u}r Physik {\bf 194}, 180 (1966)

\bibitem{Danos53} M.~Danos, S.~Geschwind, H.~Lashinsky, and A.~van~Trier, Phys.~Rev.\ {\bf 92}, 828 (1953). 

\bibitem{Takahashi} T.~Takahashi, Y.~Shibata, K.~Ishi, M.~Ikezawa, M.~Oyamada, and Y.~Kondo, Phys.~Rev.~E {\bf 62}, 8606 (2000). 

\bibitem{Gorham_formation} P.W.~Gorham, K.M.~Liewer, C.J.~Naudet, D.P.~Saltzberg, and D.R.~Williams, arXiv:astro-ph/0102435 (2001).

\bibitem{ZrelovRuzicka} V.P.~Zrelov and J.~Ru\v{z}i\v{c}ka, Czech.~J.~Phys B {\bf 39}, 368 (1989).

\bibitem{FrankTamm37} I.M.~Frank and I.E.~Tamm, Dok.~Akad.~Nauk SSSR {\bf 14}, 109 (1937).

\bibitem{GinzburgTsytovich} V.L.~Ginzburg, V.N.~Tsytovich, \emph{Transition Radiation and Transition Scattering} (Revised Ed.), Taylor \& Francis (1990).

\bibitem{KahnLerche66} F.D.~Kahn and I.~Lerche, Roal.~Soc.~London.~Proc.~Series~A {\bf 289}, 206 (1966).

\bibitem{Colgate67} S.A.~Colgate, J.Geophys.Res.\ {\bf 72}, 4869 (1967). 

\bibitem{Allan71} H.R.~Allan, Prog.~Elem.~part.~Cosm.~Ray.~Phys.\ {\bf 10}, 171 (1971).

\bibitem{Jelley} J.V.~Jelley \etal, Nature {\bf 205}, 327 (1965).

\bibitem{LOPES} H.~Falcke \etal, Nature {\bf 435}, 313 (2005).

\bibitem{CODALEMA} D.~Ardouin \etal, Astropart.~Phys.\ {\bf 26}, 341 (2006).

\bibitem{Auger_radio} T.~Huege \etal, Nucl.~Instr.~Meth. A {\bf 617}, 484 (2009).

\bibitem{MGMR} O.~Scholten, K.~Werner, and F.~Rusydi, Astropart.~Phys.\ {\bf 29}, 94 (2008).

\bibitem{EGSnrc} R.~Engel, N.N.~Kalmykov, and A.A.~Konstantinov, Int.~J.~Mod.~Phys.~A {\bf 21S1}, 65 (2006).

\bibitem{FalckeGorham03} H.~Falcke and P.~Gorham, Astropart.~Phys.\ {\bf 19}, 477 (2003).

\bibitem{HuegeFalcke03} T.~Huege and H.~Falcke, Astron.\ \& Astroph.\ {\bf 412}, 19 (2003).

\bibitem{Suprun} D.A.~Suprun, P.W.~Gorham, and J.L.~Rosner, Astropart.~Phys.\ {\bf 20}, 157 (2003).

\bibitem{REAS1} T.~Huege and H.~Falcke, Astron.\ \& Astroph.\ {\bf 430}, 779 (2005).

\bibitem{ReAires} M.A.~Duvernois, B.~Cai, and D.~Kleckner, Proc.\ 29th ICRC, Pune, India {\bf 8}, 311 (2005).

\bibitem{REAS2} T.~Huege, R.~Ulrich, and R.~Engel, Astropart.~Phys. \ {\bf 27}, 392 (2007).

\bibitem{Huege_ARENA} T.~Huege, M.~Ludwig, O.~Scholten and K.D.~deVries {\it The convergence of EAS radio emission models and a detailed comparison of REAS3 and MGMR simulations}, Presented at `Acousitc and Radio EeV Neutrino detection Activities' (ARENA), Nantes, France (2010), Nucl.~Instr.~Meth. A in press, doi:10.1016/j.nima.2010.11.041

\end{thebibliography}
\end{document}